\documentclass[aps,
pra,twocolumn,superscriptaddress,longbibliography,footinbib]{revtex4-1} 
\usepackage{amsmath,amssymb,amstext}
\usepackage{graphicx}
\usepackage[colorlinks=true,citecolor=blue,linkcolor=blue,urlcolor=blue]{hyperref}
\usepackage{physics}
\usepackage{braket}
\usepackage{float}
\usepackage{xcolor}
\usepackage{etex}

\newcommand{\bs}[1]{{\boldsymbol{#1}}}
\newcommand{\bk}{\bs{k}}
\newcommand{\bq}{\bs{q}}

\newcommand{\br}{\bs{r}}

\begin{document}
\title{Quench dynamics of a  weakly interacting disordered Bose gas in momentum space}

\author{Thibault Scoquart}
    \affiliation{Laboratoire Kastler Brossel, Sorbonne Universit\'e, CNRS, ENS-Universit\'e PSL, Coll\`ege de France; 4 Place Jussieu, 75004 Paris, France }
\author{Thomas Wellens}
    \affiliation{Physikalisches Institut, Albert-Ludwigs-Universit\"at Freiburg, Hermann-Herder-Stra\ss e 3, D-79104, Freiburg, Federal Republic of Germany}
\author{Dominique Delande}
    \affiliation{Laboratoire Kastler Brossel, Sorbonne Universit\'e, CNRS, ENS-Universit\'e PSL, Coll\`ege de France; 4 Place Jussieu, 75004 Paris, France }
\author{Nicolas Cherroret}
     \email{nicolas.cherroret@lkb.upmc.fr}
    \affiliation{Laboratoire Kastler Brossel, Sorbonne Universit\'e, CNRS, ENS-Universit\'e PSL, Coll\`ege de France; 4 Place Jussieu, 75004 Paris, France }
	
	
	\begin{abstract}
		We theoretically study the out-of-equilibrium dynamics in momentum space of a weakly interacting disordered Bose gas launched with a finite velocity. In the absence of interactions, coherent multiple scattering gives rise to a background of diffusive particles, on top of which a coherent backscattering interference emerges. We revisit this scenario in the presence of interactions, using a diagrammatic quantum transport theory. We find that the dynamics is governed by coupled kinetic equations describing the thermalization of the diffusive and coherent components of the gas. This phenomenon leads to a destruction of coherent backscattering, well described by an exponential relaxation whose rate is controlled by the particle collision time. These predictions are confirmed by numerical simulations.
\end{abstract}

\maketitle

\section{Introduction}

When perturbed from an equilibrium situation, isolated many-body systems generally experience a thermalization process and eventually return to equilibrium at sufficiently long time \cite{Polkovnikov2011}. This process arises because the interacting system serves as a ``bath'' for all its subparts, the final state being characterized by a Gibbs ensemble. A number of works have explored the formation of this thermalized state, with special attention dedicated to the dynamical emergence of a Bose condensate \cite{Lacaze2001, Connaughton2005, Sun12}.
The out-of-equilibrium dynamics leading to thermalization can also follow a rich variety of scenarios.  Those have recently raised considerable interest in the cold-atom community, where the conditions of truly isolated quantum gases can be achieved at an unprecedented  level.
Non-integrable systems, for instance, usually display an intermediate ``prethermal'' stage where the system evolves rather slowly and looks approximately thermalized  \cite{Gring2012, Eigen18}. Prethermalization can be modeled by a generalized Gibbs ensemble, characterized by a small set of parameters \cite{Langen2015}. 
The many-body dynamics may also exhibit universal scaling properties when the system is quenched through or in the vicinity of a quantum phase transition  \cite{Nicklas2015, Navon2015}, or when it is initially prepared in a far off-equilibrium state  \cite{Berges2008, Orioli2015, Erne2018, Oberthaler2018}.

Much less is known about the out-of-equilibrium dynamics of interacting \textit{disordered} systems. At the many-body level, the competition between disorder and interactions may lead to many-body localization (for recent reviews see \cite{Huse14,Alet18}), initially addressed in the context of electron conduction \cite{basko2006}. A consequence of many-body localization is the absence of thermalization. When quenched out-of-equilibrium, many-body disordered systems may or may not reach a universal thermalized state, depending on the magnitude and properties of the disorder and interactions.  In weakly interacting Bose gases, which can be described at a mean-field level with a nonlinear Schr\"odinger equation, many-body localization does not occur and thermalization is the rule. Notwithstanding the relative simplicity of this limit, the combination of weak interactions and disorder still gives rise to a number of puzzling phenomena, such as thermalization via weakly coupled localized states \cite{Mulansky2009, Basko2011} or subdiffusive spreading of wave packets \cite{pikovsky2008, skokos2009, flach2009, Cherroret2014, Cherroret2016, vakulchyk2018}.

In this article, we study the interplay between disorder and interactions in a dilute Bose gas, in the limit of weak disorder. This regime has been under the focus of a number of cold-atom  experiments probing, e.g., one-dimensional Anderson localization \cite{Billy2008, Roati2008}, coherent backscattering (CBS) \cite{Jendrzejewski2012, Labeyrie2012, Hainaut2017} as well as its control over external dephasing \cite{Josse2015, Hainaut2017}. CBS of cold atoms, in particular, was probed in an optics-like configuration where an ultracold Bose gas was initially given a finite mean velocity, and its subsequent dynamics in the presence of disorder probed in momentum space. This configuration, originally introduced in \cite{Cherroret2011}, turned also useful to explore other interference phenomena like coherent forward scattering \cite{Karpiuk2012}, to achieve an echo spectroscopy of coherent transport in disorder \cite{Micklitz2015, Josse2015} or to monitor the thermalization and dynamical formation of condensates in momentum space \cite{Cherroret2015}. 
In most of these works, disorder -- albeit weak -- was the main ingredient driving the atomic dynamics, so that interactions could be neglected in first approximation. As is well known, however, particle interactions generally affect significantly coherent transport and, in particular, coherent backscattering. This question was previously addressed in the context of nonlinear optics of continuous beams \cite{Wellens2008, Wellens2009,Geiger2012,Geiger2013} or of atom lasers \cite{Hartung2008}. A theoretical description of particle interactions in an out-of-equilibrium regime where mesoscopic effects like CBS occur is, on the other hand, still absent. It is the goal of the present work to fill this gap. 

Following \cite{Cherroret2011, Jendrzejewski2012, Josse2015, Labeyrie2012}, we consider the out-of-equilibrium dynamics of a two-dimensional, weakly interacting Bose gas initially prepared in a plane-wave state with finite velocity in a random potential. In this configuration, the momentum distribution quickly acquires a ring-shape profile associated with classical particle diffusion, with an interference, CBS peak emerging on the top. In the presence of interactions, this picture slowly evolves in time as the whole distribution thermalizes. Well before thermalization fully develops however, we observe a rather fast contrast loss of the CBS peak. To explain this phenomenon, we develop a microscopic, diagrammatic perturbation theory of coherent particle transport including both disorder and interactions. This allows us to derive coupled kinetic equations for the diffusive and coherent components of the momentum distribution. By solving the latter numerically at short time, we achieve a precise description of the time evolution of CBS in the presence of interactions.
In particular, we find that time-reversed paths responsible for CBS are quantitatively more sensitive to particle collisions than diffusive paths. This leads to a faster relaxation of the interference signal. This relaxation is well captured by an exponential decay, whose rate is controlled by the particle collision time.

The article is organized as follows. In Sec. \ref{sec:problem} we formulate the problem and illustrate it through a numerical simulation. In Sec. \ref{sec:linear_theory}, we recall the main elements of quantum transport theory in disorder when interactions are neglected. This approach is extended to the interacting regime in Secs. \ref{sec:nonlinear_theory} and \ref{sec:nonlinear_theoryCBS}, and confronted with numerical simulations in Sec. \ref{sec:comparison}. Main results are summarized in Sec. \ref{sec:conclusion}, and technical details are collected in two appendices.




\section{Momentum-space dynamics}
\label{sec:problem}

\begin{figure}[h]
	\includegraphics[scale=0.4]{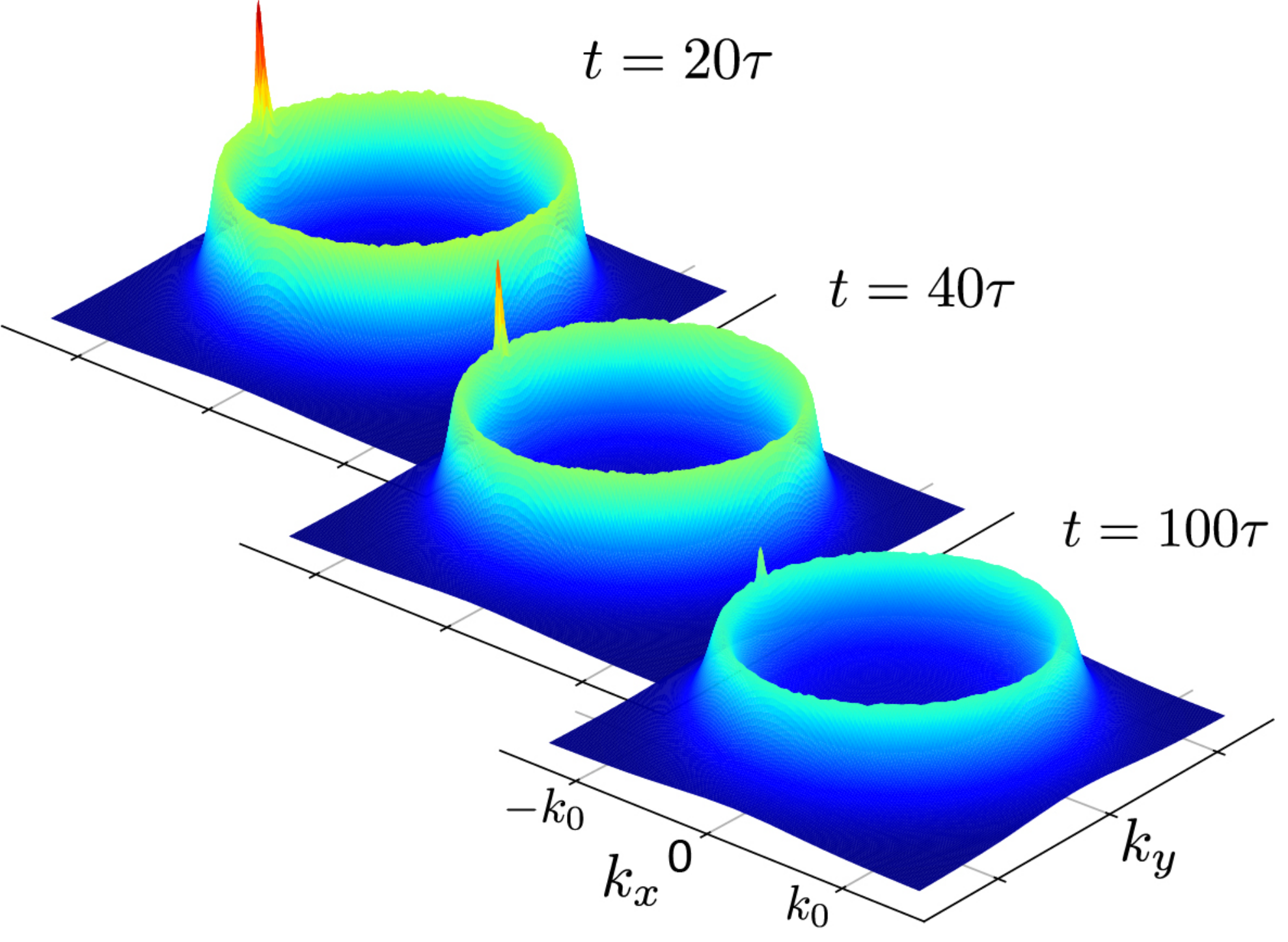}%
	\caption{Momentum distribution $\overline{|\Psi(k_x,k_y,t)|^2}$ computed numerically at three successive times  using the GPE (\ref{eq:grosspitaevskii}), starting from a plane wave of initial momentum $\bk_0=(\pi/5,0)$ in a Gaussian, uncorrelated random potential. The nonlinear term in the GPE leads to an early-time decay of the CBS peak. 
Here $g\rho_0=0.004$ and $\gamma=0.038$. Data are averaged over about $14000$ disorder realizations. 
 \label{fig:momentum_distrib}}
\end{figure}

We consider the out-of-equilibrium evolution of a $N$-particle disordered interacting Bose gas. Interactions, assumed weak, are treated at the mean-field level, on the basis of the  Gross-Pitaevskii equation (GPE)
\begin{equation}
i\partial_t\Psi(\br,t)= \left[-\frac{\boldsymbol{\nabla}^2}{2m} + V(\br)+gN|\Psi(\br,t)|^2\right]
\Psi(\br,t)
\label{eq:grosspitaevskii}
\end{equation}
for the Bose field $\Psi(\br,t)$. In Eq. (\ref{eq:grosspitaevskii}) and in the following, we set $\hbar = 1$. From now on we focus on the two-dimensional geometry, although most results of the article are valid in dimension 3 as well. $V(\mathbf{r})$ is a random potential, assumed to follow a Gaussian statistics with zero mean and no correlation :
\begin{equation}
\overline{V(\br)}=0,\ \ \ \ \  \overline{V(\br)V(\br')} = \gamma \delta(\br-\br'),
\label{eq:disorder}
\end{equation}
where $\gamma>0$ sets the disorder strength and the overbar refers to configuration averaging. 
The assumption of uncorrelated disorder does not imply any loss of generality. The mean free time (defined below) being the only relevant disorder parameter for the dynamics, the results of the paper hold as well for a short-range correlated  disorder.
To illustrate the problem we are interested in, we first study the numerical propagation of an initial plane wave $\phi(\br)\equiv\langle\br\ket{\Psi(t=0)}=1/\sqrt{V}\exp(i\bk_0\cdot\br)$ in the random potential ($V$ is the volume of the system), by computing the wave function $\Psi(\br,t)$ at different times on a discretized regular grid of 200\,$\times$\,200 sites with step $a$, for periodic boundary conditions.
The temporal propagation is performed using a split-step algorithm of time step $\Delta t$, alternating
propagations of the linear part of the GPE, $\exp[-i(-\boldsymbol{\nabla}^2/2m+V(\br))\Delta t]$, and of the nonlinear part, 
$\smash{\exp\small[-i gN \abs{\Psi(\br,t)}^2\Delta t]}$. The linear part of the evolution operator is expanded in a series of Chebyshev polynomials, as described in  \cite{Tal-Ezer84, Cheby91, roche1997conductivity,fehske2009numerical}.  From the wave function, we compute its Fourier transform
\begin{equation}
\Psi(\bk,t)\equiv 
\int{\rm d^2}\br~
e^{-i\bk\cdot\br} \Psi(\br,t),
\end{equation}
from which we infer the disorder-averaged momentum distribution, $\smash{\overline{|\Psi(\bk,t)|^2}}$, normalized according to $\smash{\int {\rm d^2}\bk/(2\pi)^2 \overline{|\Psi(\bk,t)|^2} =1}$.
The computed momentum distribution is shown in Fig. \ref{fig:momentum_distrib} at three different times. 
We choose as unit of length the discretization step $a$ in our numerics. In order for the discretization to be a good approximation to the continuous equation (\ref{eq:grosspitaevskii}), one must simply ensure that the de Broglie wavelength is significantly larger than the grid spacing, i.e. $2\pi/k_0\gg a$. In the following we typically use $k_0a=\pi/5,$ so that discretization effects are small. In Fig. \ref{fig:momentum_distrib} and in all other simulations based on Eq. (\ref{eq:grosspitaevskii}), we express momenta in units of $1/a$, lengths in $a$, times in $ma^2$ and energies in $1/(ma^2)$. The disorder amplitude, $\gamma$, is then in units of $1/(m^2a^2)$. In Fig. \ref{fig:momentum_distrib}, the three times shown are given in units of the mean free time $\tau$, i.e. the typical collision time on the scattering potential. In the Born approximation, $\tau=1/(m\gamma)$ [see Eq. (\ref{eq:scattering_time}) below]. We choose $\tau\simeq 26.3$, so that the product of $k_0$ with the mean free path $\ell\equiv k_0\tau$ is $k_0\ell\simeq 10.4$, i.e. much larger than one. This is the so-called limit of \textit{weak disorder}, where, in the absence of interactions, the momentum distribution is essentially the sum of two contributions. The first is a background of diffusive particles scattered elastically on the random potential. In Fig. \ref{fig:momentum_distrib} this contribution manifests itself as a ring of radius $k_0$. The second is a narrow interference peak centered around $\bk=-\bk_0$, the coherent backscattering (CBS) effect. CBS in this configuration was first described theoretically in \cite{Cherroret2011}, and experimentally measured with cold atoms in \cite{Jendrzejewski2012, Labeyrie2012, Josse2015}.

\begin{figure}
	\centering
	\includegraphics[scale=0.55]{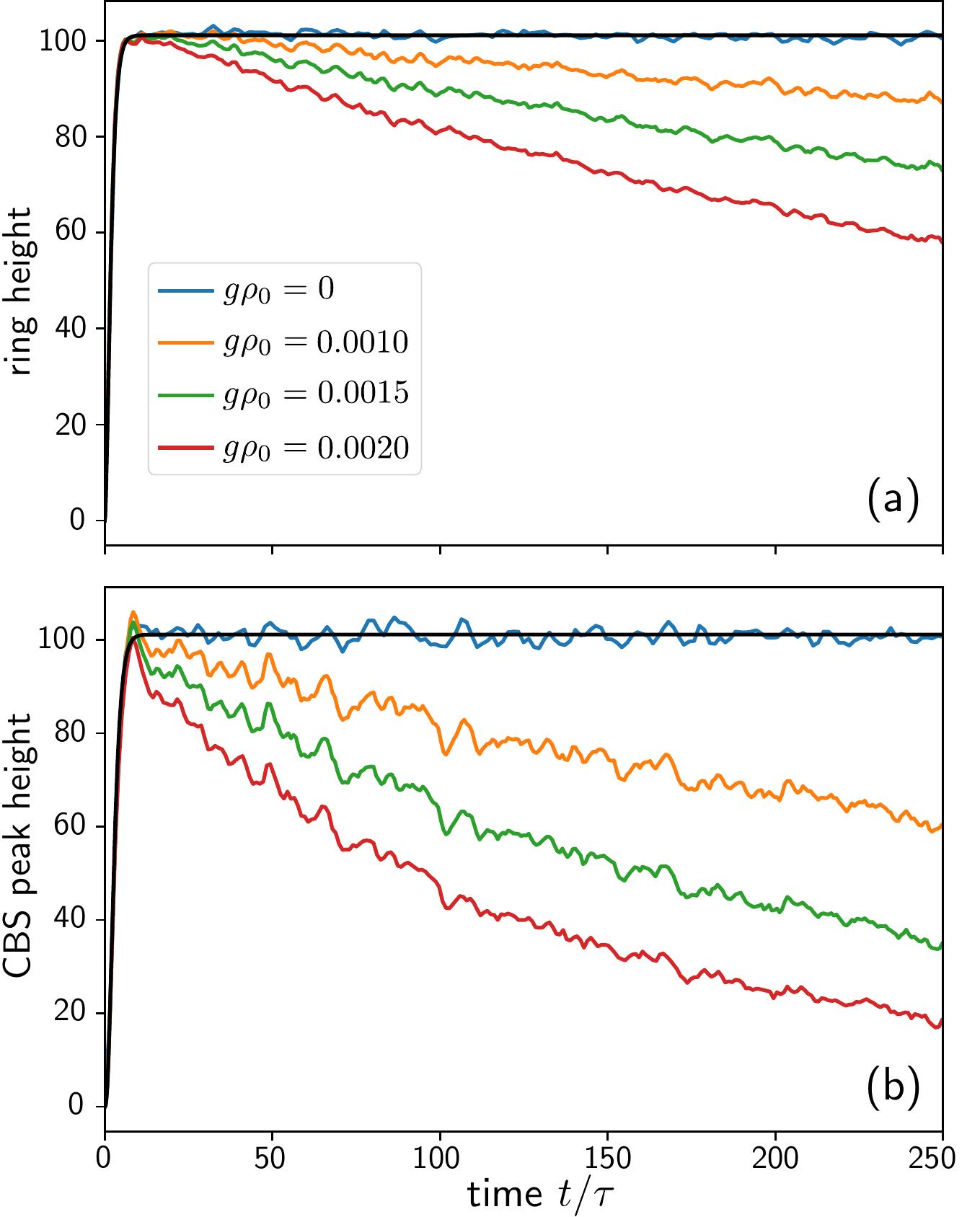}%
	\caption{Heights of the (a) diffusive ring and (b) CBS peak versus time for increasing values of $g$ from top to bottom. Data for the ring height are obtained by computing the maximum value of $\overline{|\Psi(\bk,t)|^2}$ at $(k_x,k_y)=(0,\pm k_0)$, where no CBS peak is present. Data for the CBS peak height are obtained by subtracting the momentum distribution rotated by 90$^\circ$ from $\overline{|\Psi(\bk,t)|^2}$.  Here $\bk_0=(\pi/5,0)$, $\gamma \simeq 0.0182$, and data are averaged over about 16000 disorder realizations. The solid smooth black curves are Eqs. (\ref{eq:psiD_linear}) and (\ref{eq:psiC_linear}), respectively.
They both saturate at a value slightly above 100,  which is close to the analytical estimate $ \tau/(\pi\nu_{\epsilon_0})\simeq 98.9$ for a numerically computed mean free time $\tau\simeq 51.84$ and on-shell density of state $\nu_{\epsilon_0}\simeq0.167$.
\label{fig:decay_height}}
\end{figure}

In the presence of weak interactions, a main change compared to this picture is a decay of the CBS peak amplitude at short time.
This decay appears as soon as interactions are nonzero, as  shown in the lower panel of Fig. \ref{fig:decay_height} for even weaker values of $g\rho_0$. Also shown in the upper panel is the height of the diffusive ring, which decays as well, albeit more slowly than the CBS peak. This results in a decay of the CBS contrast, well visible  in Fig. \ref{fig:momentum_distrib}.
In the simulations of Figs. \ref{fig:momentum_distrib} and \ref{fig:decay_height}, the magnitude of interactions is chosen such that the mean free path associated with the collisions on the \textit{nonlinear potential} $gN|\Psi|^2$ is larger than $\ell$ (this condition will be clarified in Sec. \ref{sec:nonlinear_theory}). In the rather small time-range of these figures, this nonlinear potential thus plays the role of a perturbation  for the dynamics, which remains mainly governed by the disorder. 

The observed  decays of the diffusive background and CBS peak constitute the early-time manifestations of a thermalization process of the whole momentum distribution. The long-time thermalization properties of the diffusive background have been previously addressed in \cite{Cherroret2015}. In the sequel of the paper, we provide a theoretical basis for the formalism used in \cite{Cherroret2015}, and go one step forward by constructing a kinetic theory  which encompasses \textit{both} the incoherent diffusive component and the CBS contribution (neglected in \cite{Cherroret2015}). Equipped with this theory, we then reproduce and explain the temporal evolutions observed in Figs. \ref{fig:momentum_distrib} and \ref{fig:decay_height}, and in particular confirm that the CBS peak is more sensitive to particle collisions than the diffusive background. The relaxation of its contrast is found to be exponential at short time, with a relaxation rate controlled by the particle collision time.

\section{Linear regime: theory}
\label{sec:linear_theory}

The theory of diffusion and CBS in momentum space has been presented in \cite{Cherroret2011}. Here we only recall the main elements required to introduce the nonlinear diagrammatic theory  in the next sections. We also adopt a slightly different point of view than in \cite{Cherroret2011}, focusing more on the \textit{energy distribution} of the particles, which plays a major role in the thermalization process at work when interactions are nonzero.

When $g=0$, the momentum distribution can be expressed as \cite{Kuhn05, Kuhn07, Cherroret2011}
\begin{equation}
\overline{|\Psi(\bk,t)|^2}=\int_{-\infty}^\infty \frac{{\rm d}\epsilon}{2\pi} \int_{-\infty}^\infty \frac{{\rm d}\omega}{2\pi}e^{-i\omega t}\mathcal{I}_{\epsilon,\omega}(\bk),
\label{eq:psik_Ieomega}
\end{equation}	
where the density kernel is defined in terms of the energy-dependent,  retarded and advanced Green's operators $\smash{G^{R/A}_\epsilon}$ and of the initial state $|\phi\rangle$ as
\begin{align}
\mathcal{I}_{\epsilon,\omega}(\bk)\equiv\int &\frac{\rm d^2\bk'}{(2\pi)^2}\frac{{\rm d}^2\bk''}{(2\pi)^2}
\overline{\langle\bk|G^R_{\epsilon+\omega/2}|\bk'\rangle\langle\bk''|G^A_{\epsilon-\omega/2}|\bk\rangle}\nonumber\\
&\times\phi(\bk')\phi^*(\bk'').
\label{eq:psik_grga}
\end{align}
In the following, we will also work with the disorder-averaged occupation number $f_\epsilon(t)$, defined as
\begin{equation}
f_\epsilon(t)\equiv\frac{1}{2\pi\nu_\epsilon}\int\frac{{\rm d}^2\bk}{(2\pi)^2}\int_{-\infty}^\infty \frac{{\rm d}\omega}{2\pi}e^{-i\omega t} \mathcal{I}_{\epsilon,\omega}(\bk),
\label{eq:energy_distrib}
\end{equation}
where $\nu_\epsilon$ is the density of states per unit volume at energy $\epsilon$. As a consequence of particle conservation, this quantity is normalized according to:
\begin{align}
\int_{-\infty}^\infty {\rm d}\epsilon\, \nu_\epsilon f_\epsilon(t)=1.
\label{eq:normalization_cond}
\end{align}
This condition identifies the product $\nu_{\epsilon} f_\epsilon(t)$ as the energy distribution of the Bose gas \footnote{Note that our definition of the occupation number $f_{\epsilon}$ does not make it dimensionless, see Eq. (\ref{eq:normalization_cond}), unlike what is more frequently encountered in the literature.
We adopt this convention because when $g\ne 0$ it makes the natural energy scale for interactions $g\rho_0\equiv gN/V$ explicitly appear in all our expressions, rather than $gN$ and $V$ separately.
A dimensionless $f_{\epsilon}$ could nevertheless be chosen by normalizing  the energy distribution $\nu_{\epsilon}f_{\epsilon}$ to $\rho_0$ rather than to unity in Eq. (\ref{eq:normalization_cond}).}.

\subsection{Diffusive background at long time}

\begin{figure}
	\includegraphics[scale=0.55]{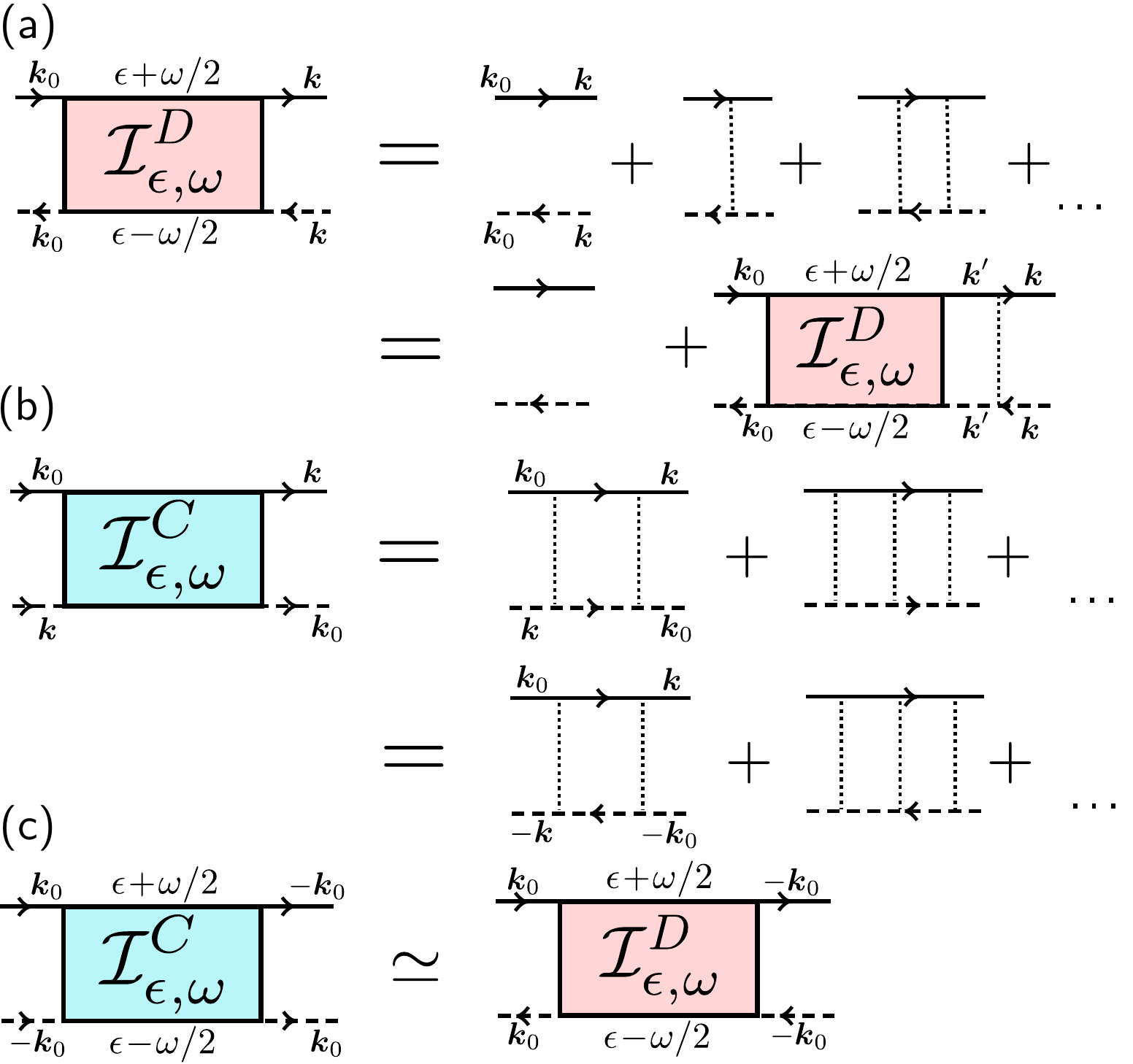}
	\caption{(a) Diagrammatic representation of the Bethe-Salpeter equation for ladder diagrams, Eq. (\ref{bethe_salpeter_linear}). 
	Upper solid lines with arrows refer  to $\langle\bk|G^R_{\epsilon+\omega/2}|\bk_0\rangle$, and lower dashed lines to $\smash{\langle\bk_0|G^A_{\epsilon-\omega/2}|\bk\rangle}$. Note that for solid (dashed) lines, arrows coincide with the (opposite of the) direction of propagation. Dotted vertical lines symbolize the correlation function in Eq. (\ref{eq:disorder}).
	(b) Series of diagrams describing the interference between time-reversed paths, responsible for CBS (``crossed diagrams''). The second equality follows from time-reversal invariance. (c) In the long-time limit $t\gg\tau$ where low scattering orders have a negligible weight, the series of ladder and crossed diagrams coincide.
	\label{bethe_salpeter_diff_linear}}
\end{figure}

From now on, we focus on the case where the initial state is a plane wave, $|\phi\rangle=|\bk_0\rangle$. 
We also  assume disorder to be weak, $k_0\ell\gg1$, so that perturbation theory can be used. The main contribution to $\mathcal{I}_{\epsilon,\omega}(\bk)$ is then given by the series of ladder diagrams (``Diffuson''), which we denote by $\mathcal{I}^D_{\epsilon,\omega}(\bk)$. 
The latter obeys the Bethe-Salpeter equation \cite{Kuhn05, Kuhn07, Cherroret2011}
\begin{align}
	\mathcal{I}^{D}_{\epsilon,\omega}(\bk) ~&=~\overline{G}^R_{\epsilon+\omega/2}(\bk)\overline{G}_{\epsilon-\omega/2}^A(\bk)
	\times(2\pi)^2\delta(\bk-\bk_0)\nonumber\\
	& +\gamma\,\overline{G}_{\epsilon+\omega/2}^R(\bk)\overline{G}_{\epsilon-\omega/2}^A(\bk)
	\int\frac{{\rm d^2}\bk'}{(2\pi)^2} \mathcal{I}^{D}_{\epsilon,\omega}(\bk'),
	\label{bethe_salpeter_linear}
\end{align}
which is shown diagrammatically in Fig. \ref{bethe_salpeter_diff_linear}a.
The average Green's function is given by
\begin{equation}
\overline{G}^{R}_{\epsilon}(\bk)=\frac{1}{\epsilon-\bk^2/2m-\Sigma(\epsilon,\bk)}.
\label{eq:GR}
\end{equation}
In the weak-disorder limit, the self energy $\Sigma(\epsilon,\bk)$ can be evaluated by perturbation theory. We restrict ourselves to the leading-order contribution provided by the Born approximation, which coincides with the Fermi golden rule:
\begin{equation}
\Im\Sigma(\epsilon,\bk)\!=\!-\pi\!\int\!\!\frac{{\rm d^2}\bq}{(2\pi)^2}B(\bk-\bq)\delta\Big(\epsilon-\frac{\bq^2}{2m}\Big)
\label{eq:self_e_born_lin}
\end{equation}
where $B(\bk)=\gamma$ is the Fourier transform of the disorder correlation function (\ref{eq:disorder}).
Equation (\ref{eq:self_e_born_lin}) defines the mean free time, 
\begin{equation}
\tau\equiv-\frac{1}{2\Im\Sigma(\epsilon,\bk)}=\frac{1}{2\pi \nu_\epsilon\gamma},
\label{eq:scattering_time}
\end{equation}
with, in the Born approximation, $\nu_\epsilon=m/(2\pi)$. In the long-time limit $t\gg\tau$, the contribution to the momentum distribution (\ref{eq:psik_Ieomega}) due to $\mathcal{I}^D_{\epsilon,\omega}$ can be obtained in the following way. First we integrate Eq.~(\ref{bethe_salpeter_linear}) over $\bk$, and expand the second term in the right-hand side for $\omega\tau\ll1$ (hydrodynamic limit). This yields
\begin{align}
\int\frac{{\rm d^2}\bk}{(2\pi)^2} \mathcal{I}^{D}_{\epsilon,\omega}(\bk)
\simeq\frac{-2\Im \overline{G}^R_{\epsilon}(\bk_0)}{-i\omega},
\label{eq:int_Ik}
\end{align}
where we used that $\overline{G}^R_{\epsilon}(\bk_0)\overline{G}^A_{\epsilon}(\bk_0)=-2\tau\Im \overline{G}^R_{\epsilon}(\bk_0)$.
We thus find, according to Eq. (\ref{eq:energy_distrib}),
\begin{align}
f_\epsilon(t)=\frac{A_\epsilon(\bk_0)}{\nu_\epsilon}\equiv f_\epsilon^{(0)}
\label{eq:spectral_gzero}
\end{align}
for the occupation number, where the superscript $(0)$ refers to the non-interacting limit and we introduced the spectral function:
\begin{equation}
A_\epsilon(\bk)\equiv -\frac{1}{\pi}\Im[\overline{G}_{\epsilon}^R(\bk)]
=\frac{1/(2\pi\tau)}{(\epsilon-\bk^2/2m)^2+1/4\tau^2}.
\end{equation}
Second, we integrate Eq.~(\ref{bethe_salpeter_linear}) over $\epsilon$ and take the Fourier transform with respect to $\omega$.
Using the result (\ref{eq:spectral_gzero}), we infer
\begin{align}
\overline{|\Psi^{D}(\bk,t)|^2} =\int_{-\infty}^\infty {\rm d}\epsilon\, 
A_\epsilon(\bk)f_\epsilon^{(0)}
\label{eq:diff_density}.
\end{align}
Eq. (\ref{eq:diff_density}) is an isotropic function of $\bk$, centered at $|\bk|=k_0$, which corresponds to the diffusive ring in Fig.~\ref{fig:momentum_distrib} \cite{Cherroret2011}. The radial profile of this ring is essentially the one of the energy distribution $\smash{\nu_\epsilon f_\epsilon^{(0)}=A_\epsilon(\bk_0)}$, which for $g=0$ coincides with the spectral function. In the absence of interactions, this energy distribution does not change in time. This is of course expected, as the only process at work is elastic multiple scattering, which does not involve any energy redistribution.

\subsection{Coherent backscattering at long time}

The CBS contribution is deduced from the diffusive one by making use of time-reversal invariance. The CBS peak stems from the interference between time-reversed multiple scattering paths, described by the diagrammatic series in Fig. \ref{bethe_salpeter_diff_linear}b (``crossed diagrams''). 
In the long-time limit $t\gg\tau$ where low scattering orders have a negligible weight, the ladder and crossed series exactly coincide at $\bk=-\bk_0$ due to time-reversal symmetry (see Figs. \ref{bethe_salpeter_diff_linear}b and c):
\begin{equation}
\mathcal{I}^C_{\epsilon,\omega}(\bk=-\bk_0)=\mathcal{I}^D_{\epsilon,\omega}(\bk=-\bk_0),
\end{equation}
such that 
\begin{align}
\overline{|\Psi^{C}(\bk=-\bk_0,t)|^2} =\int_{-\infty}^\infty {\rm d}\epsilon A_\epsilon(\bk_0)f_\epsilon^{(0)}\simeq\frac{\tau}{\pi\nu_{\epsilon_0}},
\label{eq:coop_density}
\end{align}
where $\epsilon_0\equiv\bk_0^2/(2m)$.
In the absence of interactions, the diffusive and CBS amplitudes at $-\bk_0$ thus coincide, see Eqs. (\ref{eq:diff_density}) and (\ref{eq:coop_density}), and are independent of time. The full $\bk$ dependence of the CBS profile can be calculated as well, as was done in \cite{Cherroret2011}. In the rest of the article however, we will essentially focus on its amplitude, $\smash{\overline{|\Psi^{C}(-\bk_0,t)|^2}}$.
 
\subsection{Full time evolution}

Eqs. (\ref{eq:diff_density}) and (\ref{eq:coop_density}) have been obtained in the regime of long times $t\gg\tau$, where low scattering orders can be neglected. While an exact calculation of the diffusive and CBS contributions at any time is a difficult task in general (see, for instance, \cite{Plisson2013} where this problem was tackled for a speckle potential), for the particular model of uncorrelated disorder we have found that the Bethe-Salpether equation can be exactly solved, giving: 
\begin{align}
\overline{|\Psi^{D}(-\bk_0,t)|^2}=
\rho_\text{max}
\left[1-e^{-t/\tau}\left(1+\frac{t}{\tau}\right)\right]
\label{eq:psiD_linear}
\end{align}
for the diffusive background, and
\begin{align}
\overline{|\Psi^{C}(-\bk_0,t)|^2}=\rho_\text{max}
\left[1-e^{-t/\tau}\left(1+\frac{t}{\tau}+\frac{t^2}{2\tau^2}\right)\right]
\label{eq:psiC_linear}
\end{align}
for the  amplitude of the CBS peak, with $\rho_\text{max}\equiv\int {\rm d}\epsilon A^2_\epsilon(\bk_0)/\nu_\epsilon\simeq{\tau}/{(\pi\nu_{\epsilon_0}})$.
Note that, as expected, the CBS and diffusive amplitudes coincide at long time, but not at short time $t\sim\tau$ where low scattering orders -- described by the terms within square brackets -- come into play.
Eqs. (\ref{eq:psiD_linear}) and (\ref{eq:psiC_linear}) are shown in Fig. \ref{fig:decay_height} (dashed curves) on top of the results of numerical simulations for $g=0$. The agreement is very good at all times.


\section{Interacting diffusive particles: theory}
\label{sec:nonlinear_theory}

We now turn to the case $g\ne0$. While an exact theory accounting for both interactions and disorder is a formidable task, even at the level of the GPE (\ref{eq:grosspitaevskii}), relatively simple results can be obtained when interactions are ``weak'' compared with the disorder. Indeed, in this regime the effect of interactions can be treated as a perturbation of the series of crossed and ladder diagrams. This approach was previously used in \cite{Wellens2008, Hartung2008, Wellens2009,Geiger2012} to describe the stationary coherent backscattering effect of continuous waves in finite media, and in \cite{CherroretWellens11, Schwiete2010, Schwiete2013a} to model the  dynamics of interacting wave packets  in the diffusive limit. 
The latter configuration was later extended to the localization regime in \cite{Cherroret2014, Cherroret2016}, but by  taking into account first-order corrections in $g$ only, see Sec. \ref{sec:linearg}, while neglecting second-order corrections responsible for thermalization.

In this section and the next one, we develop a  quantum transport theory describing the effect of interactions on both the diffusive and CBS signals in the \textit{dynamical} scenario of Fig. \ref{fig:momentum_distrib}. We show that, because the  average density of the Bose gas is uniform,  linear corrections  in $g$ reduce to an irrelevant shift of the mean energy, so that the  physics in momentum space is governed by second-order corrections. The latter are responsible for two coupled thermalization processes of the diffusive and CBS components.


\subsection{Weak interactions}

A treatment of the nonlinear potential in Eq. (\ref{eq:grosspitaevskii}) as a perturbation of the ladder and crossed series requires that scattering events on the nonlinear potential $g|\Psi(\br,t)|^2$
are rare compared to scattering events on the random potential $V(\br)$. In terms of time scales, this condition reads $\tau_\text{NL}\gg\tau$, where $\tau_\text{NL}$ is the particle  collision time. To estimate this quantity, we use the Fermi golden rule
\begin{equation}
\frac{1}{2\tau_\text{NL}}\!=\!\pi\int\!\!\frac{{\rm d^2}\bq}{(2\pi)^2}B_\text{NL}(\bk-\bq)\delta\Big(\epsilon-\frac{\bq^2}{2m}\Big),
\label{eq:self_e_born}
\end{equation}
where $B_\text{NL}(\bk)$ is the power spectrum of the nonlinear potential:
\begin{equation}
B_\text{NL}(\bk)\!\equiv\!\!\int\! d^2(\br\!-\!\br')g^2\!N^2\overline{|\Psi(\br,t)|^2|\Psi(\br',t)|^2}e^{i\bk(\br-\br')}.
\end{equation}
To leading approximation, the density-density correlator is not modified by interactions, and reads \cite{akkermans2007mesoscopic} 
$\overline{|\Psi(\br,t)|^2|\Psi(\br',t)|^2}=J_0(k_0|\br-\br'|)e^{-|\br-\br'|/\ell}/V^2$ in two dimensions. This gives
\begin{equation}
\tau_\text{NL}\sim \frac{\epsilon_0}{(g\rho_0)^2},
\end{equation}
where we introduced the mean particle density,
\begin{equation}
\rho_0\equiv {N/V}.
\end{equation}
By defining the  mean free path for particle collisions as $\ell_\text{NL}\equiv k_0\tau_\text{NL}/m$ we then rewrite the criteria of rare particle collisions and weak disorder as
\begin{equation}
k_0\ell_\text{NL}\gg k_0\ell\gg 1.
\label{eq:weakg_condition}
\end{equation} 
In the following, we will assume these conditions to be fulfilled. 
They imply, in particular, that the initial kinetic energy typically exceeds the interaction energy, $\epsilon_0\gg g\rho_0$. Therefore, as long as the energy distribution $\nu_\epsilon f_\epsilon(t)$ does not deviate too much from its initial value $A_\epsilon(\bk_0)$, only states belonging to the ``particle'' branch  $\epsilon\gg g\rho_0$  of the Bogoliubov spectrum are populated \cite{pitaevskii2016bose}. In other words, the low-energy, phonon-like part of the  spectrum does not play any role in the dynamics. This will be always verified in the sequel of the paper, where we  focus on the short-time evolution of the Bose gas.

\subsection{Leading-order nonlinear corrections}
\label{sec:linearg}

\begin{figure}[H]
\centering
\includegraphics[scale=0.65]{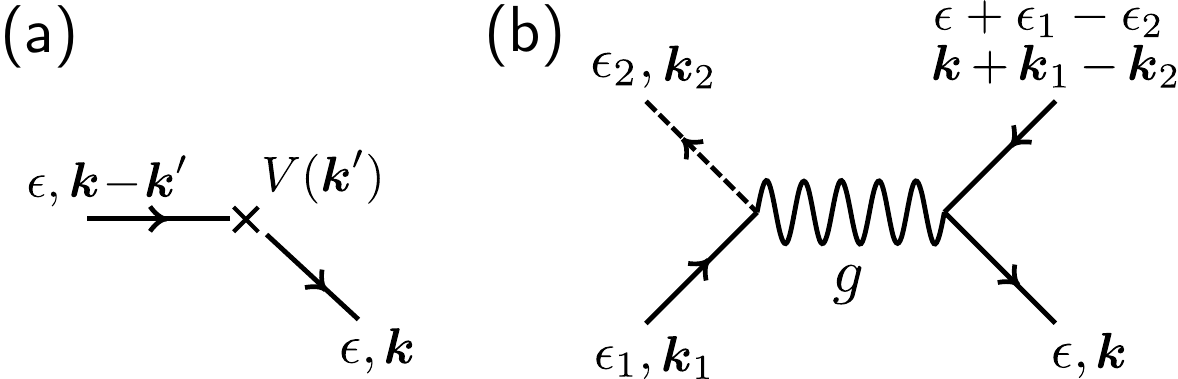}%
\caption{
The Born series obtained by iterating the Lippmann-Schwinger equation (\ref{eq:nonlinearLS}) generates terms built on scattering processes on (a) the random potential $V$ and on (b) the nonlinear potential $g|\Psi|^2$. Solid and dashed lines with arrows refer to the  free-space, retarded and advanced Green's functions, respectively. The cross refers to the random potential $V$ and the wavy line to the interaction parameter $g$. For each vertex, the lower-right line $(\epsilon,\bk)$ is the outgoing field, and integrations over $\epsilon_1$, $\epsilon_2$, $\bk'$, $\bk_1$ and $\bk_2$ are understood.
\label{feynman_rules}}
\end{figure}

When $g\ne0$, the notion of Green's function can no longer be utilized to express the momentum distribution, as we did in Eq. (\ref{eq:psik_grga}). Nevertheless, Eq. (\ref{eq:psik_Ieomega}) can still be written, with the density kernel defined as
\begin{equation}
\mathcal{I}_{\epsilon,\omega}(\bk)\equiv\overline{\Psi_{\epsilon+\omega/2}(\bk)\Psi^*_{\epsilon-\omega/2}(\bk)},
\end{equation} 
where 
\begin{equation}
\Psi_\epsilon(\bk)\equiv
\int{\rm d}^2{\br}\int_{-\infty}^\infty {\rm d}t~
e^{i\epsilon t} e^{-i\bk\cdot\br} \Psi(\br,t).
\end{equation} 
Our diagrammatic quantum transport theory in the presence of interactions is constructed from the Lippmann-Schwinger equation associated with the GPE (\ref{eq:grosspitaevskii}):
\begin{align}
\Psi_\epsilon(\bk)=\, &\phi(\bk)+G^0_\epsilon(\bk)\left[\int\frac{\rm d^2\bk'}{(2\pi)^2}V(\bk')\Psi_\epsilon(\bk-\bk')\right.\nonumber\\
&+gN\int\frac{d\epsilon_1}{2\pi}\frac{d\epsilon_2}{2\pi}
\int\frac{\rm d^2\bk_1}{(2\pi)^2}\frac{\rm d^2\bk_2}{(2\pi)^2}
\Psi_{\epsilon_1}(\bk_1)\Psi_{\epsilon_2}^*\!(\bk_2)\nonumber\\
&\times \Psi_{\epsilon-\epsilon_1+\epsilon_2}(\bk-\bk_1+\bk_2)\Big],
\label{eq:nonlinearLS}
\end{align}
where $G^0_\epsilon(\bk)=(\epsilon-\bk^2/2m+i0^+)^{-1}$ is the free-space (retarded) Green's function.
Iteration of Eq. (\ref{eq:nonlinearLS}) leads to an expansion of $\Psi_\epsilon$ in powers of $V$ and $g$ known as the Born series. In addition to the usual scattering processes on the random potential, this Born series also involves particle collisions. These two elementary processes are diagrammatically shown in Fig. \ref{feynman_rules}, together with the conservation rules for energies and momenta.
From the Born series for $\Psi_\epsilon$, one can construct a Bethe-Salpeter equation for the density kernel $\mathcal{I}_{\epsilon,\omega}(\bk)$. Insofar as particle collisions are less frequent than collisions on the random potential -- remember condition (\ref{eq:weakg_condition}) -- two distinct iterative equations for the diffusive and CBS contributions, $\mathcal{I}_{\epsilon,\omega}^D$ and $\mathcal{I}_{\epsilon,\omega}^C$, can still be identified. In this section we first focus on the Bethe-Salpeter equation for $\mathcal{I}_{\epsilon,\omega}^D$. 
\begin{figure}[h]
\includegraphics[scale=0.57]{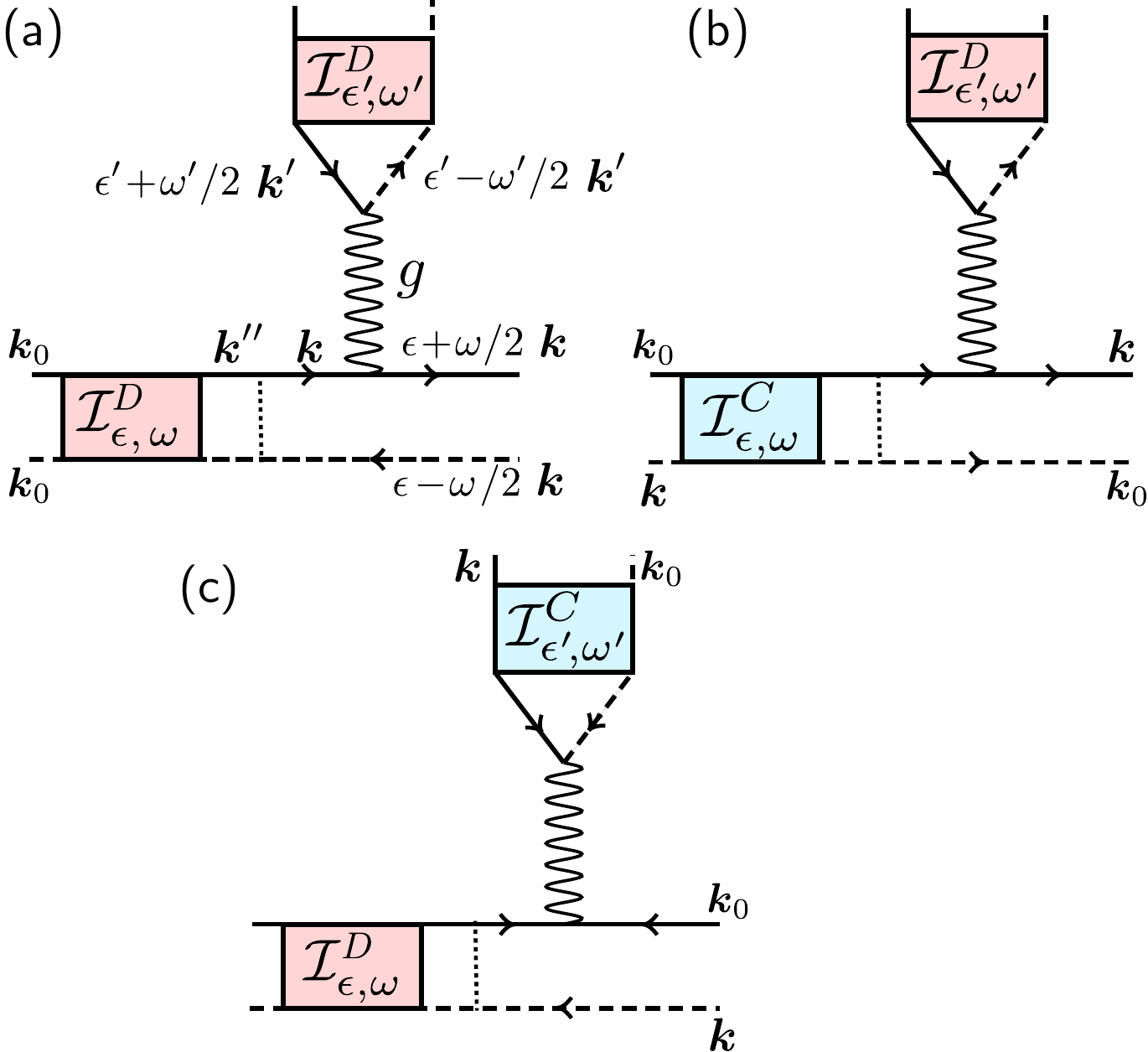}%
\caption{First-order diagrammatic corrections to the Bethe-Salpeter equation, involving one particle collision (for each diagram, one has to add the conjugate version).  (a): Correction to the Bethe-Salpeter equation for $\mathcal{I}_{\epsilon,\omega}^D$. Red boxes refer to an incoming, ladder-type sequence $\smash{\mathcal{I}^D_{\epsilon,\omega}}$, where, apart from particle collision processes, the two paths propagate along the same sequence of scatterers in the same direction. (b) and (c): Corrections to the Bethe-Salpeter equation for $\smash{\mathcal{I}_{\epsilon,\omega}^C}$. Blue boxes refer to time-reversed scattering sequences. Diagrams (a) and (b)  boil down to an irrelevant energy shift, while diagram (c) is compensated by its complex conjugate.
The solid and dashed lines symbolize $\smash{\overline{G}^R_{\epsilon}}$ and $\smash{\overline{G}^A_{\epsilon}}$, respectively, the vertical lines the correlation function in Eq. (\ref{eq:disorder}), and the wavy lines a particle collision, see Fig. \ref{feynman_rules}.
\label{fig:first_order}}
\end{figure}
The latter is obtained 
by adding extra terms to the right-hand side of the iterative equation in Fig. \ref{bethe_salpeter_diff_linear}a for $g=0$, in which the ladder sequence can be interrupted by one or several  particle collisions.

The leading-order, iterative correction to the Bethe-Salpeter equation for $\mathcal{I}^D_{\epsilon,\omega}$ is given by the diagram in Fig. \ref{fig:first_order}a (plus its complex conjugate). It reads
\begin{align}
\mathcal{I}^a_{\epsilon,\omega}(\bk)\!&=\!
2g\rho_0\gamma \!\int\!\frac{\rm d\epsilon'}{2\pi}\frac{\rm d\omega'}{2\pi}\frac{\rm d^2\bk'}{(2\pi)^2}\frac{\rm d^2\bk''}{(2\pi)^2}
\overline{G}^R_{\epsilon+\omega/2-\omega'}(\bk)\nonumber\\
&\times \overline{G}^R_{\epsilon+\omega/2}(\bk)\overline{G}^A_{\epsilon-\omega/2}(\bk)\mathcal{I}_{\epsilon',\omega'}(\bk')
\mathcal{I}_{\epsilon-\omega'\!/2,\omega-\omega'}(\bk'')\nonumber\\
&=2g\rho_0\gamma[\overline{G}^R_{\epsilon}(\bk)]^2\overline{G}^A_{\epsilon}(\bk)\int \frac{\rm d^2\bk''}{(2\pi)^2}\mathcal{I}_{\epsilon,\omega}(\bk''),
\label{eq:BSg1}
\end{align}
where the factor 2 stems from the Wick decomposition of the average $\overline{\Psi_{\epsilon_1}\Psi^*_{\epsilon_2}\Psi_{\epsilon-\epsilon_1+\epsilon_2}\Psi^*_{\epsilon}}$ arising when Eq. (\ref{eq:nonlinearLS}) is multiplied by $\Psi_\epsilon^*$, and the second equality follows from particle conservation, which imposes that
\begin{equation}
\int_{-\infty}^\infty \frac{{\rm d}\epsilon'}{2\pi}\int\frac{{\rm d}^2\bk'}{(2\pi)^2}\mathcal{I}_{\epsilon',\omega'}(\bk')=\frac{1}{-i\omega'+0^+}.
\end{equation}

From Eq. (\ref{eq:BSg1}), it is easy to  show that the contribution of the diagram in Fig. \ref{fig:first_order}a boils down to a constant energy shift $-2g\rho_0$ of $\smash{\overline{G}^R_\epsilon}$ in the linear Bethe-Salpeter equation (\ref{bethe_salpeter_linear}). Indeed, if we perform the substitution $\epsilon\to\epsilon-2g\rho_0$ in the right-hand side of Eq. (\ref{bethe_salpeter_linear}) and expand for small $g$, using 
\begin{equation}
\overline{G}^R_{\epsilon-2g\rho_0}(\bk)\simeq
\overline{G}^R_\epsilon(\bk)+2g\rho_0[\overline{G}^R_\epsilon(\bk)]^2,
\end{equation}
we get that  the right-hand side of Eq. (\ref{bethe_salpeter_linear}) is modified by an extra term which exactly coincides with Eq. (\ref{eq:BSg1}). 
In other words,  first-order nonlinear corrections to the Bethe-Salpeter equation do not quantitatively affect the diffusive dynamics. 
Note that this result is in stark contrast with the scenario where one follows the spreading of a 
wave packet in position space. In this case, diagrams of the type of Fig. \ref{fig:first_order}a were shown to significantly modify the wave-packet density distribution \cite{CherroretWellens11, Schwiete2010, Schwiete2013a}. The difference lies in the behavior of the mean density $\smash{\overline{|\Psi(\br,t)|^2}}$, which evolves in time for an initial wave packet, whereas it always remains uniform for an initial plane wave.

As explained in appendix \ref{sec:self-energy}, the energy shift obtained here can in turn be described in terms of a modification of the real part of the self energy $\Sigma(\epsilon,\bk)$ appearing in average Green's functions, Eq. (\ref{eq:GR}). In addition to this shift, there also exist first-order nonlinear corrections shifting the imaginary part of $\Sigma(\epsilon,\bk)$. These  corrections stem from correlations between the disorder and nonlinear potentials in the GPE equation (\ref{eq:grosspitaevskii}), but turn out to be very small in the weak-disorder limit. From now, we will thus neglect these self-energy corrections, and always evaluate average Green's functions using Eq. (\ref{eq:GR}), with $\Sigma(\epsilon,\bk)=-i/2\tau$.\\

\subsection{Second-order corrections: thermalization}

\begin{figure*}
	\includegraphics[scale=0.500]{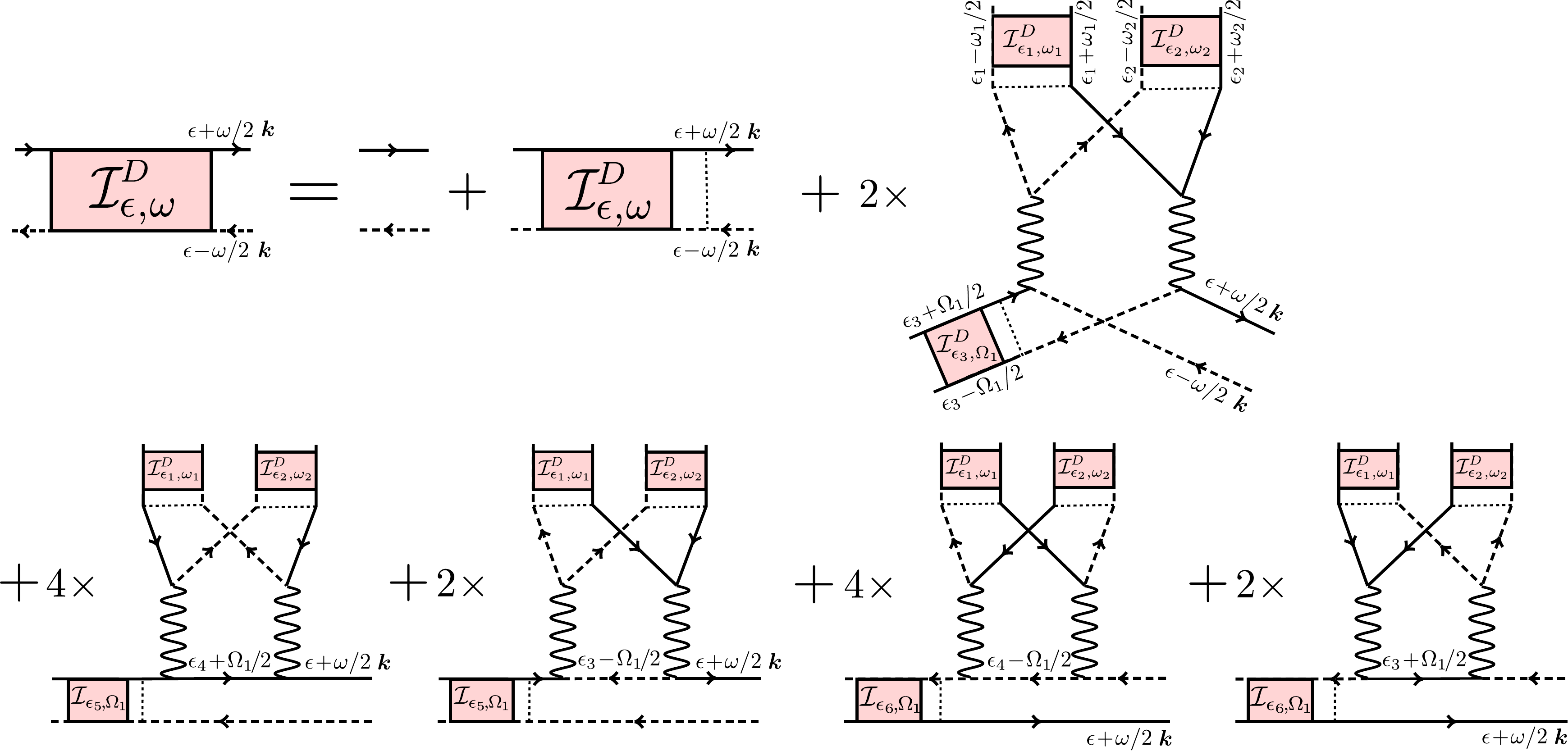}%
	\centering
	\caption{
	Diagrammatic representation of the Bethe-Salpeter equation for the diffusive contribution to the momentum distribution, $\mathcal{I}^D_{\epsilon,\omega}(\bk)$, taking into account second-order interaction corrections (first-order interaction diagrams are discarded, as explained in Sec. \ref{sec:linearg}). Symbols have the same meaning as in Fig. \ref{fig:first_order}.
The numerical prefactors account for the possible combinations of propagation lines connecting to the vertex $g$.
\label{fig:bethe_salpeter_diff}}
\end{figure*}

We now examine interaction corrections to the ladder Bethe-Salpeter equation (\ref{bethe_salpeter_linear}) that are of second order in $g$. Since each vertex $g$  is connected to four field amplitudes (see Fig.~\ref{feynman_rules}), these corrections involve six incoming field amplitudes, i.e they are proportional to the third power of the density.
Due to the condition (\ref{eq:weakg_condition}), we also know that at least 
 one disorder scattering event occurs before every particle collision event.
Since
the disorder scattering events are described by ladder diagrams (for weak disorder), we group the
six incoming arrows into three incoming ladder sequences $\mathcal{I}^D_{\epsilon_i,\omega_i}$, each of them originating from different
disorder scattering events. Analyzing all possible non-trivial ways (i.e. those which do not reduce to a mere energy shift) of connecting the incoming arrows to the $g$ vertices, we arrive
at the diagrams shown in Fig.~\ref{fig:bethe_salpeter_diff}. The corresponding Bethe-Salpeter equation reads:
\begin{widetext}
	\begin{align}
	&\mathcal{I}_{\epsilon,\omega}(\bk)  = \overline{G}^R_{\epsilon+\omega/2}(\bk)\overline{G}^A_{\epsilon-\omega/2}(\bk)\!~\Bigg[~
	(2\pi)^2\delta(\bk-\bk_0) + \gamma\int\frac{{\rm d^2}\bk'}{(2\pi)^2}  \mathcal{I}_{\epsilon,\omega}(\bk')\Biggr.\nonumber\\
	& +  
	\!{(g\rho_0\gamma)^2}\!
	\Big[\!\prod_{i=1,2}\!
	\int\!\frac{{\rm d}\epsilon_i{\rm d}\omega_i}{(2\pi)^4}
	\frac{{\rm d^2}\bk_i{\rm d^2}\bk_i'}{(2\pi)^8}
	\mathcal{I}_{\epsilon_i,\omega_i}(\bk_i')
	\overline{G}^R_{\epsilon_i+\omega_i/2}(\bk_i)
	\overline{G}^A_{\epsilon_i-\omega_i/2}(\bk_i)\Big]
\! \Biggl\{\!2\gamma\overline{G}^R_{\epsilon_3+\Omega_1/2}(\bk_3)\overline{G}^A_{\epsilon_3-\Omega_1/2}(\bk_3)\biggr. \!\int\!\frac{{\rm d^2}\bk_3'}{(2\pi)^2}  \mathcal{I}_{\epsilon_3,\Omega_1}(\bk_3')\nonumber\\ 
	&\hspace{3cm} + \gamma\Bigl(4\overline{G}^R_{\epsilon_4+\Omega_1/2}(\bk_4)+2 \overline{G}^A_{\epsilon_3-\Omega_1/2}(\bk_3)\Bigr)\nonumber 
	\overline{G}^R_{\epsilon+\Omega_2}(\bk)\int\frac{{\rm d^2}\bk'}{(2\pi)^2}  \mathcal{I}_{\epsilon_5,\Omega_1}(\bk')\nonumber\\
	& \hspace{3cm}+ \gamma\Bigl(4\overline{G}^A_{\epsilon_4-\Omega_1/2}(\bk_4)+2 \overline{G}^R_{\epsilon_3+\Omega_1/2}(\bk_3)\Bigr)
	\Biggl.\biggl. 
	\overline{G}^A_{\epsilon-\Omega_2}(\bk)
	\int\frac{{\rm d^2}\bk'}{(2\pi)^2}  \mathcal{I}_{\epsilon_6,\Omega_1}(\bk')\Biggr\}\Bigg],
	\label{eq:bethesalpeterD}
	\end{align}
\end{widetext}
where we defined $\epsilon_3=\epsilon_1+\epsilon_2-\epsilon$, $\epsilon_4=\epsilon+\epsilon_1-\epsilon_2$, $\epsilon_5=\epsilon-(\omega_1+\omega_2)/2$, $\epsilon_6=\epsilon+(\omega_1+\omega_2)/2$, $\Omega_1=\omega-\omega_1-\omega_2$, $\Omega_2=\omega/2-\omega_1-\omega_2$ for energies, and $\bk_3=\bk_1+\bk_2-\bk$, $\bk_4=\bk+\bk_1-\bk_2$ for momenta. 

A closed equation for the distribution $f_\epsilon(t)$ -- remember the definition (\ref{eq:energy_distrib}) --  can be obtained by integrating Eq. (\ref{eq:bethesalpeterD}) over $\bk$ and taking the Fourier transform with respect to $\omega$, in the hydrodynamic limit $\omega\tau,\omega_i\tau\ll1$ ($i=1,2$). 
The details of the calculation are presented in Appendix \ref{app:kinetic_eq} for clarity. They lead to the kinetic equation
\begin{align}
&\partial_t f_\epsilon=\int_{\epsilon_1,\epsilon_2,\epsilon_3\geq0}\!\!\!\!\!\!\!\!\!\!\!\!\!\!\!\!\!\!{\rm d}\epsilon_1{\rm d}\epsilon_2~W(\epsilon,\epsilon_1,\epsilon_2)\times\nonumber\\
&\Bigl[(f_{\epsilon}+f_{\epsilon_1+\epsilon_2-\epsilon})f_{\epsilon_1}f_{\epsilon_2}-f_{\epsilon}f_{\epsilon_1+\epsilon_2-\epsilon}(f_{\epsilon_1}+f_{\epsilon_2})\Bigr]
\label{eq:kinetic_eqD}
\end{align}
where  we recall that $\epsilon_3\equiv\epsilon_1+\epsilon_2-\epsilon$. The interaction kernel is given, in two dimensions, by
\begin{align}
W(\epsilon;\epsilon_1,\epsilon_2)=\frac{m^3(g\rho_0)^2}{2\pi^4\nu_\epsilon}\frac{K\left(\frac{ 2\sqrt[4]{\epsilon\epsilon_1\epsilon_2\epsilon_3}}{\sqrt{\epsilon_1\epsilon_2}+\sqrt{\epsilon\epsilon_3}}\right)}{\sqrt{\epsilon_1\epsilon_2}+\sqrt{\epsilon\epsilon_3}},
\label{eq:kinetic_kernel}
\end{align}
where $K$ is the complete elliptic integral of the first kind. The kinetic equation should be complemented by an initial condition, which is here provided by the ``coherent mode'', i.e. the first term in the right-hand side of Eq. (\ref{eq:bethesalpeterD}). The latter gives (see Appendix \ref{app:kinetic_eq}):
\begin{equation}
f_\epsilon(t=0)=\frac{A_\epsilon(\bk_0)}{\nu_\epsilon}\equiv f_\epsilon^{(0)},
\label{eq:initial_cond}
\end{equation}
which is nothing but the occupation number for $g=0$, Eq. (\ref{eq:spectral_gzero}). The fact that the non-interacting value of $f_\epsilon$ plays the role of the initial condition for the interacting problem is due to our assumption that particle collisions are less frequent than scattering events on the disorder, Eq. (\ref{eq:weakg_condition}). Indeed, in this regime the diffusive ring first establishes, and only then do interactions come into play. 

Once $f_\epsilon(t)$ is known for $g\ne0$,
 the diffusive contribution to the momentum distribution follows by integrating Eq. (\ref{eq:bethesalpeterD}) over $\epsilon$ and taking the Fourier transform with respect to $\omega$. 
As the effect of interactions is already included in the second term of the right-hand side via $\mathcal{I}_{\epsilon,\omega}(\bk)$, the third term is typically of order $g^4$ and can be neglected. In the hydrodynamic regime $\omega\tau\ll1$, this finally gives:
\begin{align}
\overline{|\Psi^{D}(\bk,t)|^2} =\int_{-\infty}^\infty {\rm d}\epsilon\, 
A_\epsilon(\bk)f_\epsilon(t).
\label{eq:diff_density_g}
\end{align}
This formula differs from its
 non-interacting counterpart, Eq. (\ref{eq:diff_density}), in that the distribution $f_\epsilon(t)$ is no longer constant in time, leading to an evolution of the diffusive background.
Equations (\ref{eq:kinetic_eqD}) and (\ref{eq:diff_density_g})   have been used, in particular, in \cite{Cherroret2015}, to qualitatively discuss the emergence of a Bose condensate at very long time, but without  microscopic justification. We note that the kinetic equation (\ref{eq:kinetic_eqD}) has also been derived in \cite{Schwiete2013b} by means of a non-equilibrium classical field theory in the presence of disorder.

By multiplying the kinetic equation (\ref{eq:kinetic_eqD}) by $\nu_\epsilon$ and integrating over $\epsilon$, we readily obtain $\partial_{t}\int{\rm d}\epsilon\, \nu_\epsilon\, f_\epsilon(t)=0$. This implies that $\smash{\int{\rm d}\epsilon\, \nu_\epsilon\, f_\epsilon(t)=\int{\rm d}\epsilon\, \nu_\epsilon f_\epsilon^{(0)}=1}$, which is nothing but the normalization condition (\ref{eq:normalization_cond}). 
It follows that the diffusive contribution (\ref{eq:diff_density_g}) is normalized, $\smash{\int {\rm d}^2\bk/(2\pi)^2\overline{|\Psi^{D}(\bk,t)|^2}=1}$, very much like in the non-interacting limit \cite{akkermans2007mesoscopic}.

The attentive reader will notice that Eq. (\ref{eq:kinetic_eqD}) in fact coincides with the \textit{free-space} Boltzmann kinetic equation for Bose gases  in the limit of large occupation numbers \cite{Griffin09}. Indeed, the kernel (\ref{eq:kinetic_kernel}) is independent of any disorder parameter (in particular,  the diffusion coefficient or even the mean free path do not appear). This result is different from the case of electrons in low-temperature disordered conductors, where the diffusive motion strengthens the effect of interactions \cite{Altshuler1985}. 
This difference stems from the mechanism of dynamical screening of electron-electron interactions, which is absent for low-temperature bosons. \cite{Ashcroft76}
For weakly interacting diffusive bosons, disorder thus only manifests itself through the spectral function, involved both in the initial condition $f_\epsilon(0)$ and in Eq. (\ref{eq:diff_density_g}). The situation would of course change at stronger disorder or in the localization  regime 
\cite{Basko2011}.

\section{CBS of interacting particles: theory}
\label{sec:nonlinear_theoryCBS}

\subsection{Leading-order nonlinear corrections}

\begin{figure}
	\centering
	\includegraphics[scale=0.5]{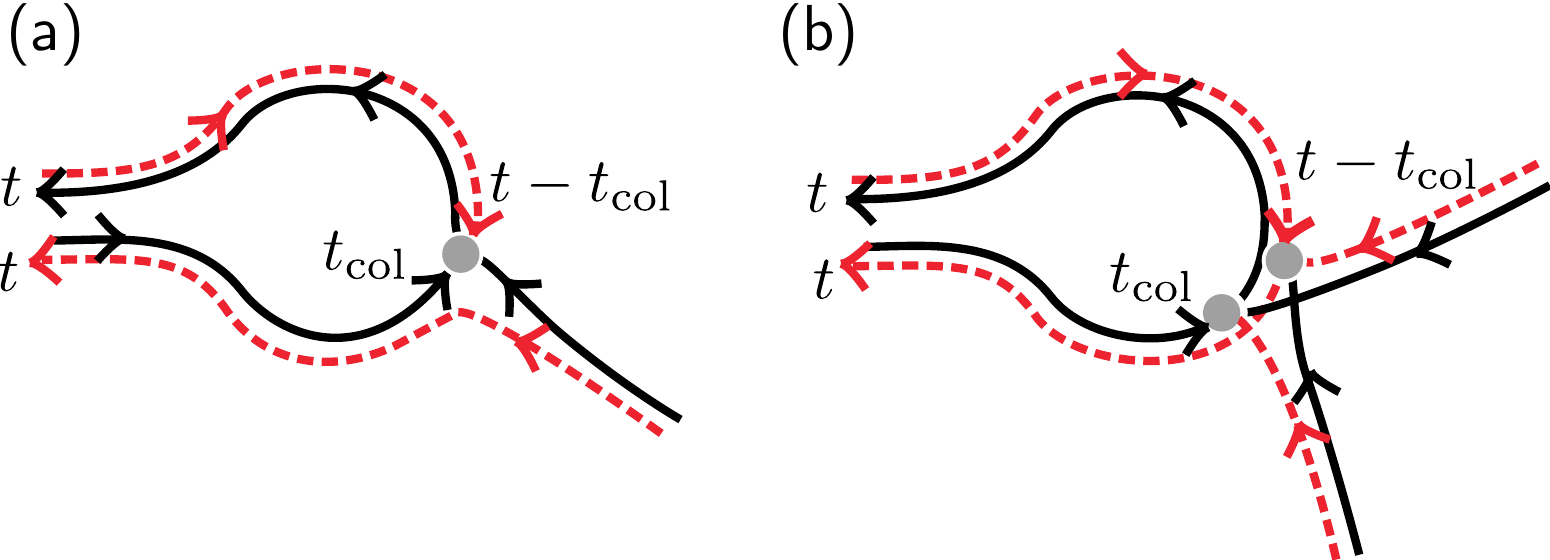}%
	\caption{
(a) Example of interference sequence generated by the diagram in Fig. \ref{fig:first_order}c (for a better visualization we momentarily change the definition of arrows, which here always indicate the direction of propagation). The direct path, starting at $t=0$, undergoes a particle collision at $t_\text{col}$, and the time reversed-path at $t-t_\text{coll}$. Since the collision is local in time, we must have $t_\text{coll}=t/2$.
(b) Example of interference sequence between time-reversed paths involving two collisions. As for diagram (a), the collision processes involve both the direct and the reversed paths. This  imposes them to occur almost simultaneously, at $t_\text{coll}\simeq t/2$, which is very unlikely. This diagram is therefore negligible.
 \label{fig:neglected_diagrams}}
\end{figure}

We now come to the central part of our work and examine the effect of interactions on the series of time-reversed paths, responsible for coherent backscattering. As for the diffusive background, we first address the first-order nonlinear corrections to the Bethe-Salpeter equation of Fig. \ref{bethe_salpeter_diff_linear}b.
These corrections are displayed in Fig. \ref{fig:first_order}b and \ref{fig:first_order}c. The diagram \ref{fig:first_order}b has the very same property as its incoherent counterpart \ref{fig:first_order}a: it can be recast as an energy shift $-2g\rho_0$ of the linear Green's function, and thus does not play any role in the dynamics. The building block \ref{fig:first_order}c, on the other hand, turns out to cancel with its conjugate counterpart.
At this stage, an important comment is in order. In the stationary scenario considered in \cite{Wellens2008, Hartung2008, Wellens2009}, it was shown that specific concatenations of the diagram \ref{fig:first_order}c was leading to a dephasing between the reversed amplitudes, which could even change the sign of the coherent backscattering cone. It turns out, however, that in the present dynamical setup these combinations have a negligible weight. To see this, we show in a more visual fashion in Fig. \ref{fig:neglected_diagrams}a one example of interference sequence between time-reversed paths built from diagram \ref{fig:first_order}c.
The peculiarity of this sequence is that both the direct (solid) path and its time-reversed (dashed) partner are involved in the particle collision process. If the direct path undergoes the collision at a certain time $t_\text{coll}$, and thus the time-reversed path at time $t-t_\text{coll}$, the temporal locality  of the collision imposes that  $t_\text{coll}=t-t_\text{coll}$, i.e. that $t_\text{coll}=t/2$. In other words, the collision must occur at a very specific time (more precisely, within a time window of width $\tau$, centered around $t/2$). 
Within such a short time window, and given the condition (\ref{eq:weakg_condition}), it is highly unlikely that two (or more) collisions  occur. 
Any concatenation of diagrams \ref{fig:first_order}c can thus be safely neglected here, and an examination of second-order corrections is again required.

\subsection{Second-order corrections}

\begin{figure}
\includegraphics[scale=0.5]{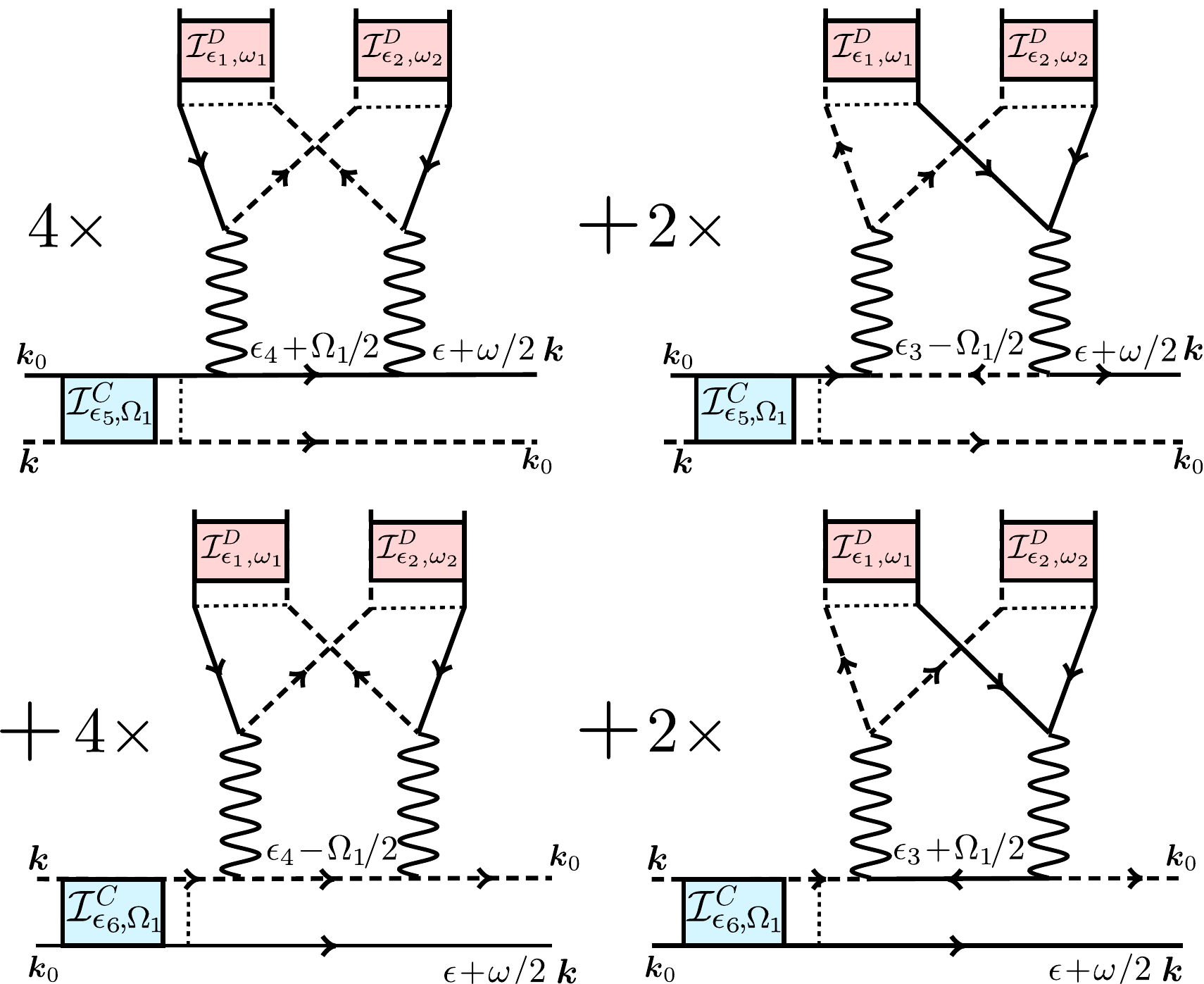}
\caption{Second-order corrections to the Bethe-Salpeter equation for the CBS contribution to the momentum distribution, $\mathcal{I}^C_{\epsilon,\omega}(\bk)$.
Symbols have the same meaning as in Fig. \ref{fig:first_order}. Recall that the two wave paths involved in the blue boxes propagate in opposite directions (the incoming and outgoing momenta $\bk_0$ and $\bk$ are explicitly indicated for clarity).
The numerical prefactors account for the possible combinations of propagation lines connecting to the vertex $g$. 
\label{bethe_salpeter_coop}}.
\end{figure}

All non-negligible second-order corrections to the Bethe-Salpeter equation for time-reversed sequences are depicted in Fig. \ref{bethe_salpeter_coop}.
Note that, as compared to the diffusive corrections in Fig. \ref{fig:bethe_salpeter_diff}, there are only four different topologies, and not five. 
These topologies are the same as those of the  four last diffusive diagrams in Fig. \ref{fig:bethe_salpeter_diff}. 
It turns out, indeed, that the interference counterpart 
of the upper-right diagram in Fig. \ref{fig:bethe_salpeter_diff} is negligible in the long-time limit. This can be understood from Fig. \ref{fig:neglected_diagrams}b, which shows the interference  sequence that would correspond to the upper-right diagram in Fig. \ref{fig:bethe_salpeter_diff} in which an amplitude is time-reversed. Because the two collision processes involve both the direct and the reversed paths, they must occur almost simultaneously, at $t_\text{coll}\simeq t/2$, which is extremely unlikely given the rarity of particle collisions assumed here. This type of diagram is thus  negligible.

\begin{figure}
	\centering 
	\includegraphics[scale=0.55]{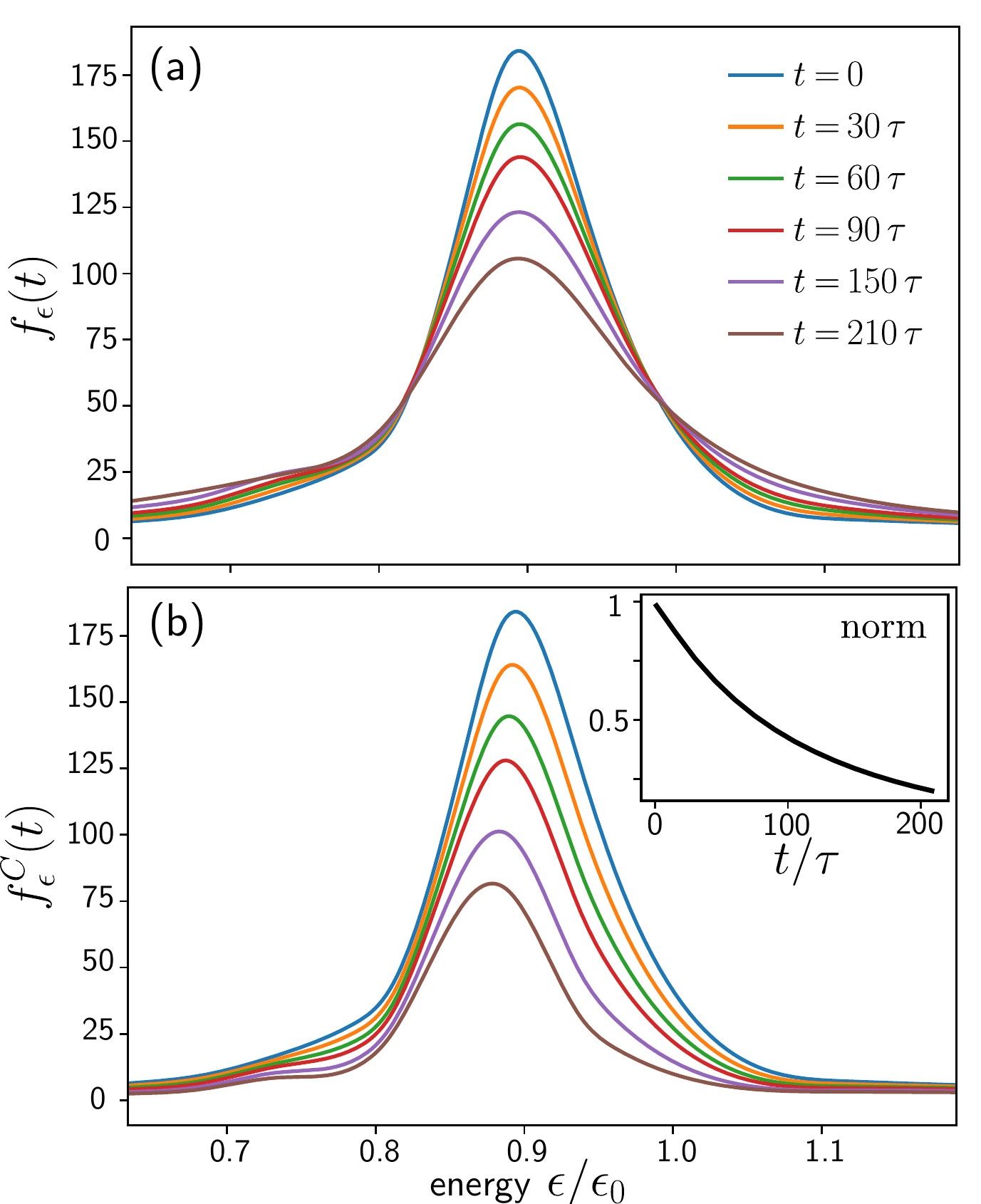}%
	\caption{
Distributions (a) $f_\epsilon(t)$ and (b) $f_\epsilon^C(t)$ at increasing times from top to bottom, obtained by solving the kinetic equations (\ref{eq:kinetic_eqD}) and (\ref{eq:kinetic_eqC}) for $\bk_0=(\pi/5,0)$, $g\rho_0=0.002$, and $\gamma=0.0182$, with the initial condition (\ref{eq:initial_cond}) evaluated numerically for $g=0$. The different evolutions reflect the asymmetry of the kinetic equations: as time grows the distributions $f_\epsilon$ broaden, whereas the $f_\epsilon^C$ flatten out.
 This is emphasized by the inset, which shows that the norm of the energy distribution \smash{$\int_{-\infty}^\infty {\rm d}\epsilon\, \nu_\epsilon f_\epsilon^C(t)$} decays in time (whereas $\int_{-\infty}^\infty {\rm d}\epsilon\, \nu_\epsilon f_\epsilon(t)$ is unity at all times).
 \label{fig:fet_feCt}}
\end{figure}

The Bethe-Salpeter equation for $\mathcal{I}^C_{\epsilon,\omega}$ is similar to Eq. (\ref{eq:bethesalpeterD}) except for the missing diagram. The latter is responsible for the term $\propto f_{\epsilon_1+\epsilon_2-\epsilon} f_{\epsilon_1} f_{\epsilon_2}$ in Eq.~(\ref{eq:kinetic_eqD}). Since this term is now absent, the $f_\epsilon(t)$ present in each of the other terms can be factored out. We thus obtain, for the ``coherent'' occupation number
\begin{equation}
f^C_\epsilon(t)\equiv\frac{1}{2\pi\nu_\epsilon}\int\frac{{\rm d}^2\bk}{(2\pi)^2}\int_{-\infty}^\infty \frac{{\rm d}\omega}{2\pi}e^{-i\omega t} \mathcal{I}^C_{\epsilon,\omega}(\bk),
\label{eq:energy_distribC}
\end{equation}
the kinetic equation
\begin{align}
\partial_t f^C_\epsilon=f^{C}_{\epsilon}&\int_{\epsilon_1,\epsilon_2,\epsilon_3\geq0}\!\!\!\!\!\!\!\!\!\!\!\!\!\!\!\!\!\!{\rm d}\epsilon_1{\rm d}\epsilon_2~W(\epsilon,\epsilon_1,\epsilon_2)\nonumber\\
&\times\Bigl[f_{\epsilon_1}f_{\epsilon_2}-f_{\epsilon_1+\epsilon_2-\epsilon}(f_{\epsilon_1}+f_{\epsilon_2})\Bigr],
\label{eq:kinetic_eqC}
\end{align}
where the kernel is still given by Eq. (\ref{eq:kinetic_kernel}), and the initial condition is again set by the non-interacting limit, $\smash{f^C_\epsilon(t=0)=f^{(0)}_\epsilon}$. The amplitude of the coherent backscattering peak then follows from:
\begin{equation}
\overline{|\Psi^C(-\bk_0,t)|^2}  =  \int_{-\infty}^\infty{\rm d}\epsilon\, A_\epsilon(\bk_0)f^{C}_\epsilon(t).
\label{eq:mom_distrib_coop}
\end{equation}
The asymmetry between the kinetic equations for $f^C_\epsilon$ and $f_\epsilon$ explains the different  
 dynamic evolution of  the CBS peak and diffusive background observed in numerical simulations, Fig. \ref{fig:decay_height} 
 \footnote{As is well known, adding CBS to the diffusive contribution leads to a breakdown of normalization when $g=0$. This is also the case when $g\ne 0$, as seen from the fact that $\int {\rm d}\epsilon\, \nu_\epsilon f_\epsilon^C(t)\ne 0$. Restoring normalization requires a fine account of weak localization corrections \cite{Knothe13}, which is beyond the scope of this paper.}. This asymmetry is a major difference with the non-interacting regime. 
 We show in Fig. \ref{fig:fet_feCt} the evolution of $f_\epsilon(t)$ and $f_\epsilon^C(t)$ at short time, obtained by solving the kinetic equations (\ref{eq:kinetic_eqD}) and (\ref{eq:kinetic_eqC}) with the initial condition (\ref{eq:initial_cond}) evaluated numerically for $g=0$. To find the latter, we  have computed the spectral function and the density of states numerically as explained in \cite{Trappe15, Prat16}. The thermalization mechanism is well visible in the upper graph: the distribution $f_\epsilon(t)$ broadens as time grows. A different behavior is observed for $f_\epsilon^C(t)$, which does not  broaden but rather flattens out. This phenomenon 
is already visible at the level of Eq. (\ref{eq:kinetic_eqC}), in which $f^C_\epsilon(t)$ factorizes out of the collision integral. It is also emphasized by the inset of Fig. \ref{fig:fet_feCt}, which shows that the norm $\int{\rm d}\epsilon\, \nu_\epsilon f_\epsilon^C(t)$ decays in time. This is in contrast with $\int {\rm d}\epsilon\, \nu_\epsilon f_\epsilon(t)$, which is unity at all times; see the discussion following Eq. (\ref{eq:diff_density_g}). 
This difference implies that time-reversed paths are more sensitive to particle collisions that diffusive paths, as  we will show more quantitatively in the next section.

Note, in passing, that the energy corresponding to the maximum of the distributions in Fig. \ref{fig:fet_feCt} lies always slightly below $\epsilon=\epsilon_0$. This effect, also seen in experiments \cite{Volchkov18}, is mainly due to the real part of the self-energy (see appendix \ref{sec:self-energy}).

\section{Comparison with numerical simulations}
\label{sec:comparison}

In order to test our theoretical approach, we now confront the predictions of the last section to numerical simulations of plane-wave propagation based on the GPE (\ref{eq:grosspitaevskii}). For this purpose, we integrate numerically the collision integrals  in the kinetic equations (\ref{eq:kinetic_eqD}) and (\ref{eq:kinetic_eqC}).

\subsection{Diffusive ring and CBS peak amplitudes}

\begin{figure}
	\centering
	\includegraphics[scale=0.62]{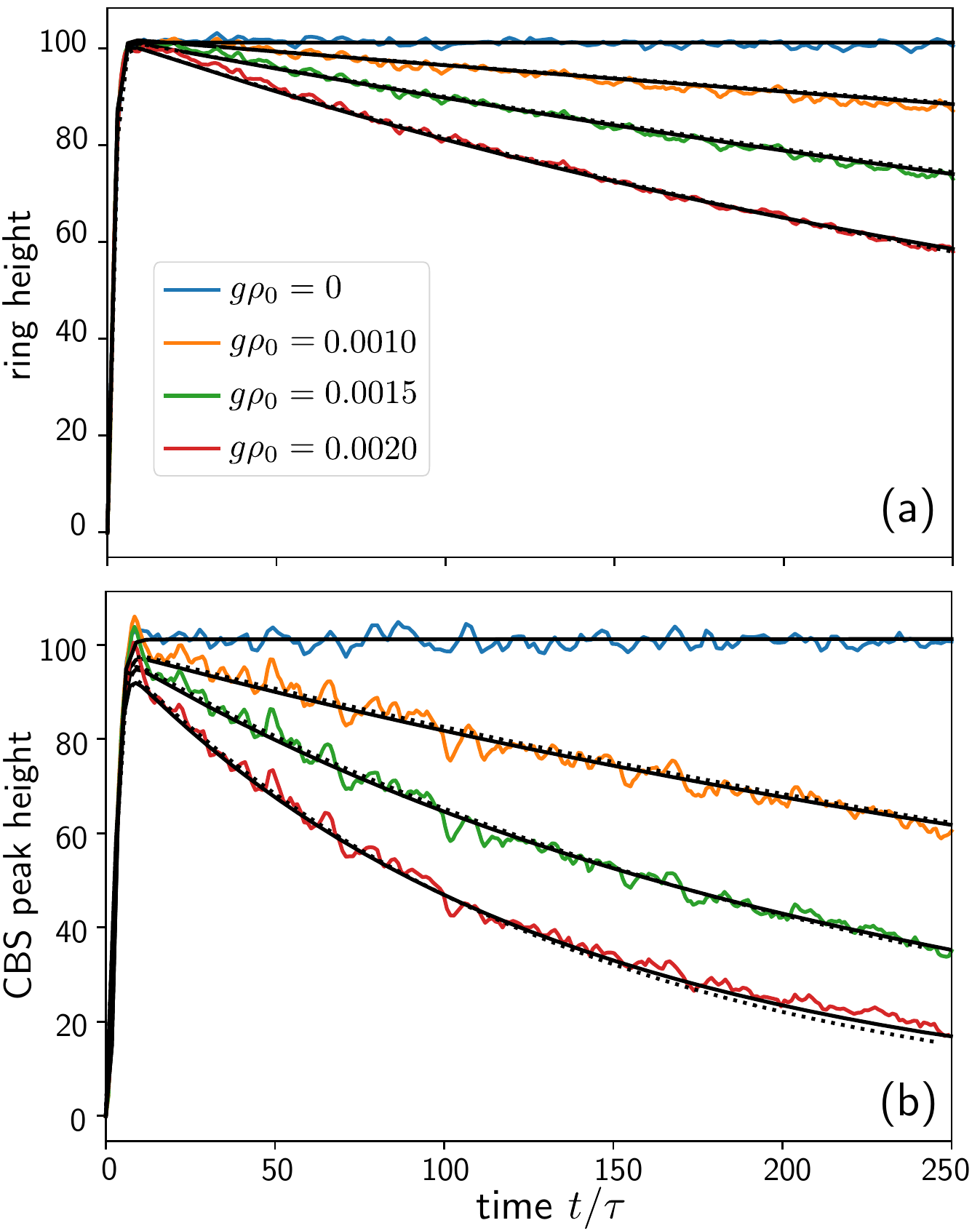}%
	\caption{
			Heights of the (a) diffusive ring and (b) CBS peak  versus time for increasing values of $g\rho_0$ from top to bottom (these are the same simulation data as in Fig. \ref{fig:decay_height}). Solid smooth black curves are fits to the theory, Eqs. (\ref{eq:diff_density_g}) and (\ref{eq:mom_distrib_coop}), and dotted curves are the approximate laws (\ref{eq:psiC_nonlinear}) and (\ref{eq:psiD_nonlinear}). The same value $g_\text{fit}\simeq 1.1g$ (resp. $g_\text{fit}\simeq 1.3g$) was used for all the diffusive (resp. CBS) curves. Solid curves at $g=0$ are obtained from  Eqs. (\ref{eq:psiD_linear}) and (\ref{eq:psiC_linear}).
\label{decay_height_theory}}
\end{figure}

In Fig. \ref{decay_height_theory}, we reproduce the simulation results of Fig. \ref{fig:decay_height}  for the heights of the diffusive ring and CBS peak. For $g\ne0$, we fit them with Eqs. (\ref{eq:diff_density_g}) and (\ref{eq:mom_distrib_coop}), with $f_\epsilon$ and $f_\epsilon^C$ computed from the kinetic equations  (\ref{eq:kinetic_eqD}) and (\ref{eq:kinetic_eqC}), using $g\rho_0$ as a fit parameter. To describe the times $t\leq \tau$, we multiply the right-hand side of Eqs. (\ref{eq:diff_density_g}) and (\ref{eq:mom_distrib_coop}) by the same short-time corrections as in the linear case [terms within the square brackets in Eqs. (\ref{eq:psiD_linear}) and (\ref{eq:psiC_linear})]. This is a very good approximation in the regime $\tau_\text{NL}\gg\tau$ considered here.
As seen in Fig. \ref{decay_height_theory}, the agreement between theory and simulations is excellent at all times. We note, however, that for the ring height the fitted values of $g\rho_0$ differ by $\sim10\%$ compared to those chosen in numerical simulations, and for the CBS height by $\sim 30\%$. One possible reason for this difference might come from the numerical uncertainties in the resolution of the collision integrals. Evaluating the latter indeed turns particularly challenging in two dimensions, where the kernel $W$ exhibits a number of logarithmic singularities over the integration domain. This discrepancy could also stem from higher-order interaction contributions that renormalize the interaction strength $g$ and not taken into account in the present work. Further investigation would however be required to clarify this point.

\subsection{Decay rates}

An important information one may extract from the time-dependent evolutions in Fig. \ref{decay_height_theory} are the characteristic time scales $\smash{\tau_\text{NL}^{D,C}}$ governing the decay of the diffusive ring and of the CBS amplitude. 
Theoretically, these characteristic times can be obtained from a short-time expansion of the  solution of the kinetic equations (\ref{eq:kinetic_eqD}) and (\ref{eq:kinetic_eqC}), 
$f_\epsilon^{D,C}(t)\simeq f_\epsilon(t=0)-\alpha_{D,C}(\epsilon)t+\mathcal{O}(t^2)$, from which we obtain, using Eq. (\ref{eq:diff_density_g}) and (\ref{eq:mom_distrib_coop}):
\begin{equation}
\overline{|\Psi^{D,C}(-\bk_0,t)|^2}\simeq\frac{\tau}{\pi\nu_{\epsilon_0}}
\left[1-\frac{t}{\tau_\text{NL}^{D,C}}+\mathcal{O}(t^2)\right],
\end{equation}
where 
\begin{equation}
(\tau_\text{NL}^{D,C})^{-1}\equiv
\frac{\pi\nu_{\epsilon_0}}{\tau}
\int {\rm d}\epsilon\,  A_\epsilon(\bk_0) \alpha_{D,C}(\epsilon).
\label{eq:tauNLDC}
\end{equation}
Here we used -- see Eq. (\ref{eq:coop_density}) -- that $\int {\rm d}\epsilon A_\epsilon(\bk_0)f_\epsilon(t=0)=\tau/(\pi\nu_{\epsilon_0})$.
By inserting the Taylor expansions for $f_\epsilon$ and $f_\epsilon^C$ in the kinetic equations, we find the functions $\alpha_{D,C}(\epsilon)$ numerically and, from Eq. (\ref{eq:tauNLDC}), infer the sought out time scales by numerical integration over $\epsilon$. This leads to
\begin{equation}
(\tau_\text{NL}^{D,C})^{-1}=\frac{(g\rho_0)^2}{\epsilon_0}\beta_{D,C},
\label{eq:tauDC_result}
\end{equation}
where $\beta_D\simeq2.27$ and $\beta_C\simeq 7.17$ are numerical prefactors which include the adjustment of the interaction strength used for the fits in Fig. \ref{decay_height_theory}. 
\begin{figure}
	\includegraphics[scale=0.65]{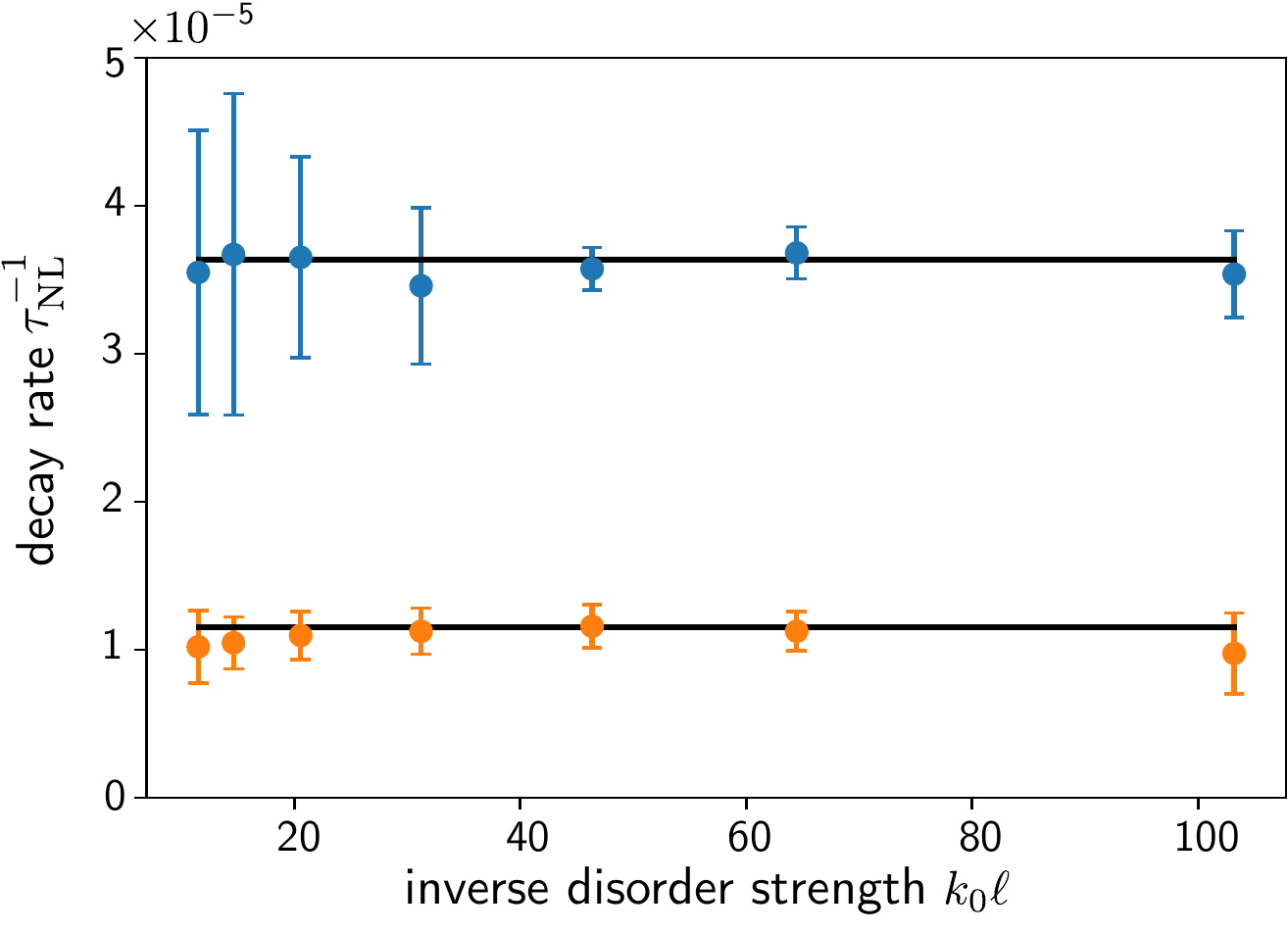}%
	\caption{
	Decay rates $(\tau_\text{NL}^{D,C})^{-1}$ for the diffusive background (lower orange dots) and CBS (upper blue dots) amplitudes, at fixed $g\rho_0=0.001$. Dots are obtained from numerical simulations of the Gross-Pitaevskii equation, by computing the slope of the simulation curves in Fig. \ref{decay_height_theory} for several values of $\gamma$ (error bars originate from the fitting of the slopes by a straight line). From right to left: $\gamma=$ 0.0036, 0.0056, 0.0081, 0.0121, 0.0182, 0.0256, 0.0324, 0.0506. Data are displayed as a function of the dimensionless disorder parameter $k_0\ell$. They confirm the theoretical predictions (\ref{eq:tauDC_result}), shown as solid lines, and in particular the independence of the decay rates on the disorder strength.
\label{fig:slopes}}
\end{figure}
The characteristic times governing the decay of the diffusive background and CBS peak are therefore both proportional to the particle collision time, which we previously introduced in Eq. (\ref{eq:self_e_born}). This is quite a natural result, but it should be noted that the decay time for CBS is approximately three times smaller than the decay time for the diffusive background. In other words, the CBS peak is way more sensitive to interactions, as clearly seen in Fig. \ref{decay_height_theory}.

To confirm these results, we also compare the theoretical prediction (\ref{eq:tauDC_result}) to the decay rates extracted from numerical simulations based on the GPE. For this purpose, we compute numerically the CBS and diffusive signals versus time for several values of the disorder amplitude $\gamma$. We then extract the slope of these curves within a narrow time window following the curve maxima (located near $t=15\tau$ in Fig. \ref{decay_height_theory}). The results are shown in Fig. \ref{fig:slopes} as a function of the disorder parameter $k_0\ell$ (lower points are obtained from the decay of the diffusive ring and upper points from the decay of the CBS peak). The theoretical predictions (\ref{eq:tauDC_result}) are shown on the same graph, and they match very well the simulations. These results confirm, in particular, the independence of the collision time on the disorder. By computing numerically the slopes for several values of $g$, we have also verified  the $g^2$ dependence of $\smash{(\tau_\text{NL}^{D,C})^{-1}}$.

While we have not been able to find an exact analytical prediction for the whole time decay of the CBS peak, a simple, approximate expression can be inferred from the kinetic equation for $f_\epsilon^C$, Eq. (\ref{eq:kinetic_eqC}). Indeed, since the  time-dependence of the diffusive background -- encoded in the $f_{\epsilon_i}$ functions in the right-hand side of  Eq. (\ref{eq:kinetic_eqC}) -- is rather slow, in first approximation the occupation number $f_\epsilon^C(t)$ decays exponentially. This suggests the simple form
\begin{align}
&\overline{|\Psi^{C}(-\bk_0,t)|^2}\simeq
\frac{\tau}{\pi\nu_{\epsilon_0}}\exp\left(-{t}/{\tau_\text{NL}^C}\right)\label{eq:psiC_nonlinear}\\
&\times\left[1-\exp(-{t}/{\tau})\left(1+{t}/{\tau}+{t^2}/{2\tau^2}\right)\right]\nonumber
\end{align}
for the CBS peak amplitude (the second line is the short-time evolution, which, we recall, is not modified by interactions). Eq. (\ref{eq:psiC_nonlinear}) is shown in Fig. \ref{decay_height_theory} (dashed curves of the lower plot) on top of the exact solutions of the kinetic equation, and turns out to be a rather good approximation. 
We also noticed that a similar formula describes well the decay of the diffusive ring, provided  $\tau_\text{NL}^D$ is substituted for $\tau_\text{NL}^C$ and the short-time terms are modified according to Eq. (\ref{eq:psiD_linear}):
 \begin{align}
 &\overline{\strut|\Psi^{D}(-\bk_0,t)|^2}\simeq
 \frac{\tau}{\pi\nu_{\epsilon_0}}\exp\left(-{t}/{\tau_\text{NL}^D}\right)\label{eq:psiD_nonlinear}\\
 &\times\left[1-\exp(-{t}/{\tau})\left(1+{t}/{\tau}\right)\right].\nonumber
 \end{align}
Eq. (\ref{eq:psiD_nonlinear}) is also shown in Fig. \ref{decay_height_theory} (dashed curves of the upper plot) on top of the exact solutions of the kinetic equation.
As a consequence of Eqs. (\ref{eq:psiC_nonlinear}) and (\ref{eq:psiD_nonlinear}), the contrast of the CBS peak decays exponentially at long time as 
\begin{align}
\frac{\overline{|\Psi^{C}(-\bk_0,t)|^2}}{\overline{\strut|\Psi^{D}(-\bk_0,t)|^2}}\simeq \exp(-t/\tau_\phi),
\end{align}
with a relaxation rate  
$\smash{\tau_\phi^{-1}\equiv(\tau_\text{NL}^\text{C})^{-1}-(\tau_\text{NL}^\text{D})^{-1}}\simeq 5(g\rho_0)^2/\epsilon_0^2$ controlled by the particle collision time.

\section{Conclusion}
\label{sec:conclusion}

In this article, we have constructed a microscopic diagrammatic theory describing the out-of-equilibrium evolution  of a weakly interacting disordered Bose gas in momentum space. Assuming weak disorder and rare particle collisions, we have derived coupled kinetic equations for the two main physical processes at work in this regime,  particle diffusion and coherent backscattering. Our approach has revealed a noticeable asymmetry in the kinetic equations for these two contributions, implying a faster decay of the CBS peak at short time and thus a loss of CBS contrast. 
We have shown that this contrast loss is very well described by an exponential relaxation, whose rate is governed by the particle-particle collision time. This phenomenon ressembles the smoothing of the weak localization correction to the conductivity due to the finite electronic coherence time associated with electron-electron interactions  in disordered conductors, but it here occurs in a non-equilibrium context.
Natural extensions of our work concern the role of interactions in the localization regime, where the phenomenon of coherent forward scattering shows up in momentum space \cite{Karpiuk2012}, or the properties of the Kosterlitz-Thouless transition expected in the equilibrium state reached at long time \cite{Carleo2013}. Related open questions also include the exploration of the opposite regime of a disorder weaker than interactions -- where the phononic part of the boson spectrum is expected to come into play --, the possible existence of nonthermal fixed points in the presence of disorder, or the out-of-equilibrium dynamics in the many-body regime.

\section*{Acknowledgments}

The authors are grateful to Tony Prat and Denis Basko for discussions.
NC acknowledges financial support from the Agence Nationale de la Recherche (grant ANR-19-CE30-0028-01 CONFOCAL).

\appendix

\section{Self-energy}
\label{sec:self-energy}

\begin{figure}
\includegraphics[scale=0.74]{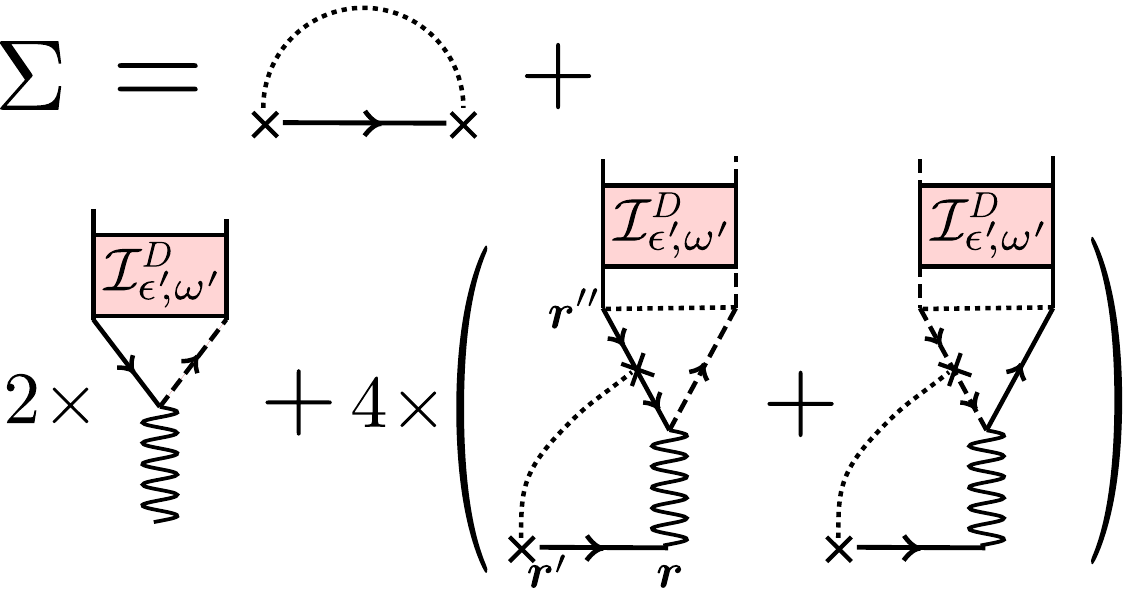}%
\centering
\caption{Leading-order contributions to an expansion of the self energy when $g\ne0$. Symbols have the same meaning as in Fig. \ref{fig:first_order}.
\label{fig:self-energy}}
\end{figure}

As discussed in Sec. \ref{sec:linearg}, leading-order nonlinear corrections to the Bethe-Salpeter equation boil down to an irrelevant energy shift. This shift can thus be alternatively described in terms of a self-energy correction, linear in $g$. The corresponding expansion of $\Sigma(\epsilon,\bk)$ is displayed in Fig. \ref{fig:self-energy}.
The first diagram is the usual Born approximation for the disorder potential,
\begin{align}
\Sigma^{(0)}(\epsilon,\bk)=\int\frac{{\rm d}^2\bq}{(2\pi)^2}B(\bk-\bq)G^{(0)}(\bq),
\label{eq:sigma0}
\end{align}
with $G^0_\epsilon(\bk)=(\epsilon-\bk^2/2m+i0^+)^{-1}$ the free-space Green's function. 
The imaginary part of Eq. (\ref{eq:sigma0})  defines the scattering mean free time (\ref{eq:scattering_time}). 
As visible in Fig. \ref{fig:fet_feCt}, the average energy of bosons in the disorder potential is also shifted, which is due to the real part  of the self-energy. In two dimensions, the latter is divergent within the Born approximation (\ref{eq:sigma0}), but it can be regularized via more refined approximations which will not be discussed here. The second self-energy diagram is the nonlinear correction corresponding to the density diagram in Fig. \ref{fig:first_order}a. It is purely real, simply given by
\begin{equation}
\Sigma^{(1)}(\epsilon,\bk)=2g\rho_0,
\label{eq:sigma1}
\end{equation}
and indeed describes a shift of the energy by $-2g\rho_0$  in the Green's function (\ref{eq:GR}).
At first order in $g$ however, another type of self-energy diagram comes into play. The latter follows from the observation that the two random potentials $V(\br)$ and $g|\Psi(\br,t)|^2$ in the GPE (\ref{eq:grosspitaevskii}) may be \textit{correlated}. This defines an analogous version of Eq. (\ref{eq:sigma0}),
\begin{equation}
\Sigma^{(2)}(\epsilon,\bk)=\int\frac{{\rm d}^2\bq}{(2\pi)^2}B_\text{NL}^{(2)}(\bk-\bq)G_\epsilon^{(0)}(\bq),
\label{eq:sigma2}
\end{equation}
with the power spectrum
\begin{equation}
B_\text{NL}^{(2)}(\bk)\!\equiv\!\!\int\! d^2(\br\!-\!\br')4gN\overline{|\Psi(\br,t)|^2V(\br')}e^{i\bk(\br-\br')}.
\end{equation}
In the factor $4$ added, one factor $2$ stems from the two possibilities to pair the incoming fields into ladder intensities [same factor as in Eq. (\ref{eq:sigma1})], and another factor 2 counts the other combination $\smash{\overline{|\Psi(\br',t)|^2V(\br)}}$. The self-energy diagrams corresponding to Eq. (\ref{eq:sigma2}) are shown in the lower-right part of Fig. \ref{fig:self-energy}.

The self energy $\Sigma^{(2)}$ corresponds to a screening effect where fluctuations of the random potential are
smoothed by the nonlinearity, and is well known in the Thomas-Fermi regime of strong interactions \cite{Sanchez-Palencia06}.
The hybrid correlator is conveniently expressed in position space as follows:
\begin{align}
&\overline{|\Psi(\br,t)|^2V(\br')}=\gamma^2\int{\rm d}^2 \br''\int\frac{{\rm d}\epsilon'}{2\pi}\frac{{\rm d}\omega'}{2\pi}e^{-i\omega' t}\mathcal{I}_{\epsilon',\omega'}(\br'')\nonumber\\
&\times[G^{(0)*}_{\epsilon'}(\br''\!-\!\br)G^{(0)}_{\epsilon'}(\br''\!-\!\br')G^{(0)}_{\epsilon'}(\br'\!-\!\br)+\text{c.c.}].
\end{align}
This simplifies to:
\begin{align}
&\overline{|\Psi(\br,t)|^2V(\br')}=\nonumber\\
&\ \ \ \ \ \frac{i \gamma}{2\pi\nu_{\epsilon_0}V}
[G^{(0)}_{\epsilon_0}(\br-\br')^2-G^{(0)*}_{\epsilon_0}(\br-\br')^2],
\end{align}
where we invoked particle conservation and used that the energy remains peaked around $\epsilon\simeq\epsilon_0\equiv\bk_0^2/2m$ at short time.
The self energy (\ref{eq:sigma2}) can  then be rewritten as
\begin{align}
\Sigma^{(2)}(\epsilon,\bk)\simeq& \frac{4g\rho_0\gamma i}{2\pi\nu_{\epsilon_0}}\int {\rm d}^2\br\,
 e^{-i\bk\cdot\br}G^{(0)}_\epsilon(\br)\nonumber\\
&\times
[G^{(0)}_{\epsilon_0}(\br)^2-G^{(0)*}_{\epsilon_0}(\br)^2].
\end{align}
Evaluating the integral on-shell, i.e. for $\epsilon=\epsilon_0$ and $\bk=\bk_0$, we obtain the following estimate for the imaginary part of $\Sigma^{(2)}$:
\begin{align}
\Im\Sigma^{(2)}(\epsilon_0,\bk_0)\sim\frac{g\rho_0}{k_0\ell}\ll g\rho_0.
\end{align}
In the weak-disorder regime considered throughout the paper, this contribution to the self energy is thus negligible. A similar decay with $k_0\ell$ is also expected for the real part of $\Sigma^{(2)}$, though its precise form requires a regularization beyond the Born approximation, as for $\Sigma^{(0)}$.

\section{Derivation of the kinetic equation}
\label{app:kinetic_eq}

To obtain the kinetic equation (\ref{eq:kinetic_eqD}), we proceed as follows. First, we integrate the Bethe-Salpeter equation (\ref{eq:bethesalpeterD}) over $\bk$. This leads to a closed equation for the quantity:
\begin{equation}
\mathcal{I}_{\epsilon,\omega}\equiv\int \frac{\rm d^2\bk}{(2\pi)^2}\mathcal{I}_{\epsilon,\omega}(\bk).
\end{equation}
In the hydrodynamic regime $\omega\tau\ll1$ (long times), we also simplify the second term in the right-hand side of Eq. (\ref{eq:bethesalpeterD}), using that
\begin{equation}
\gamma\int\frac{\rm d^2\bk}{(2\pi)^2}\overline{G}^R_{\epsilon+\omega/2}(\bk)\overline{G}^A_{\epsilon-\omega/2}(\bk)
\simeq1+i\omega\tau.
\end{equation}
Within the same limit, we also set all $\omega$, $\omega_1$ and $\omega_2$ to zero in the frequency arguments of the Green's functions in the interaction term of the right-hand side. To perform the remaining integrals over $\bk$, $\bk_1$ and $\bk_2$, finally, we use several times the identity:
\begin{equation}
\overline{G}^R_{\epsilon}(\bk)\overline{G}^A_{\epsilon}(\bk)\!=\!\frac{i}{2\pi\nu_\epsilon\gamma}
\!\int\! {\rm d}^2\br e^{-i\bk\cdot\br}
[\overline{G}^R_{\epsilon}(\br)\!-\!\overline{G}^A_{\epsilon}(\br)]
\end{equation}
and systematically neglect products of the type $\smash{\overline{G}^R_{\epsilon}(\br)\overline{G}^R_{\epsilon}(\br)}$ and $\smash{\overline{G}^A_{\epsilon}(\br)\overline{G}^A_{\epsilon}(\br)}$, which give contributions smaller by a factor $1/(k_0\ell)\ll1$.
Eq. (\ref{eq:bethesalpeterD}) becomes:
\begin{widetext}
\begin{align}
-i\omega\mathcal{I}_{\epsilon,\omega}&\!=\!i[\overline{G}^R_{\epsilon}(\bk_0)-\overline{G}^A_{\epsilon}(\bk_0)]
+\frac{(g\rho_0)^2}{(2\pi)^3}
\!\prod_{i=1,2}\int\!\frac{{\rm d}\epsilon_i{\rm d}\omega_i}{(2\pi)^2}
\frac{\mathcal{I}_{\epsilon_i,\omega_i}}{\nu_{\epsilon_i}}
\!\int\!{\rm d}^2\br
[\overline{G}^R_{\epsilon_1}(\br)-\overline{G}^A_{\epsilon_1}(\br)]
[\overline{G}^R_{\epsilon_2}(\br)-\overline{G}^A_{\epsilon_2}(\br)]
[\overline{G}^R_{\epsilon}(\br)-\overline{G}^A_{\epsilon}(\br)]\nonumber\\
&\times
\left\{
2\frac{\mathcal{I}_{\epsilon_1+\epsilon_2-\epsilon,\omega-\omega_1-\omega_2}}{\nu_{\epsilon_1+\epsilon_2-\epsilon}}
[\overline{G}^R_{\epsilon_1+\epsilon_2-\epsilon}(\br)-\overline{G}^A_{\epsilon_1+\epsilon_2-\epsilon}(\br)]
+2\frac{\mathcal{I}_{\epsilon,\omega-\omega_1-\omega_2}}{\nu_{\epsilon}}[\overline{G}^R_{\epsilon_1+\epsilon_2-\epsilon}(\br)-\overline{G}^A_{\epsilon_1+\epsilon_2-\epsilon}(\br)]\right.\nonumber\\
&\left.-4\frac{\mathcal{I}_{\epsilon,\omega-\omega_1-\omega_2}}{\nu_{\epsilon}}[\overline{G}^R_{\epsilon+\epsilon_1-\epsilon_2}(\br)-\overline{G}^A_{\epsilon+\epsilon_1-\epsilon_2}(\br)]
\right\}.
\label{eq:BS_derivation}
\end{align}
\end{widetext}
We then Fourier transform with respect to $\omega$ and use that
\begin{equation}
\frac{1}{2\pi\nu_\epsilon}\int \frac{{\rm d}\omega}{2\pi}e^{-i\omega t}\mathcal{I}_{\epsilon,\omega}\equiv f_\epsilon(t).
\end{equation}
This finally yields
\begin{align}
&\partial_t f_\epsilon=\delta(t)\frac{A_\epsilon(\bk_0)}{\nu_\epsilon}+
\int
{\rm d}\epsilon_1{\rm d}\epsilon_2~W(\epsilon,\epsilon_1,\epsilon_2)\times\nonumber\\
&\Bigl[(f_{\epsilon}+f_{\epsilon_1+\epsilon_2-\epsilon})f_{\epsilon_1}f_{\epsilon_2}-f_{\epsilon}f_{\epsilon_1+\epsilon_2-\epsilon}(f_{\epsilon_1}+f_{\epsilon_2})\Bigr],
\end{align}
which is the kinetic equation (\ref{eq:kinetic_eqD}). The Dirac-delta term originates from the first term in the right-hand side of Eq. (\ref{eq:BS_derivation}) (coherent mode), and sets the initial condition, $f_{\epsilon}(t=0)=A_\epsilon(\bk_0)/\nu_\epsilon$. The integration range of $\epsilon_1$ and $\epsilon_2$ covers all energies allowed by the densities of states $\nu_{\epsilon_1}$, $\nu_{\epsilon_2}$ and $\nu_{\epsilon_1+\epsilon_2-\epsilon}$  contained in the definition of the occupation numbers. At weak disorder, the density of states coincides, at leading order, with the free-space one, which imposes $\epsilon_1$, $\epsilon_2$, $\epsilon_3\geq0$.
The kernel derived from Eq. (\ref{eq:BS_derivation}) is given by:
\begin{align}
W(\epsilon,\epsilon_1,\epsilon_2)=&\frac{(g\rho_0)^2}{4\pi^3\nu_\epsilon}\int{\rm d}^2\br\,
[\overline{G}^R_{\epsilon}(\br)-\overline{G}^A_{\epsilon}(\br)]
\nonumber\\
&\times[\overline{G}^R_{\epsilon_1}(\br)-\overline{G}^A_{\epsilon_1}(\br)]
[\overline{G}^R_{\epsilon_2}(\br)-\overline{G}^A_{\epsilon_2}(\br)]\nonumber\\
&\times[\overline{G}^R_{\epsilon_1+\epsilon_2-\epsilon}(\br)-\overline{G}^A_{\epsilon_1+\epsilon_2-\epsilon}(\br)].
\end{align}
The expression (\ref{eq:kinetic_kernel}) follows by computing the integral over $\br$, keeping only the leading-order contribution in $k_0\ell\gg1$ \cite{Watson66}.


\begin{thebibliography}{64}%
\makeatletter
\providecommand \@ifxundefined [1]{%
 \@ifx{#1\undefined}
}%
\providecommand \@ifnum [1]{%
 \ifnum #1\expandafter \@firstoftwo
 \else \expandafter \@secondoftwo
 \fi
}%
\providecommand \@ifx [1]{%
 \ifx #1\expandafter \@firstoftwo
 \else \expandafter \@secondoftwo
 \fi
}%
\providecommand \natexlab [1]{#1}%
\providecommand \enquote  [1]{``#1''}%
\providecommand \bibnamefont  [1]{#1}%
\providecommand \bibfnamefont [1]{#1}%
\providecommand \citenamefont [1]{#1}%
\providecommand \href@noop [0]{\@secondoftwo}%
\providecommand \href [0]{\begingroup \@sanitize@url \@href}%
\providecommand \@href[1]{\@@startlink{#1}\@@href}%
\providecommand \@@href[1]{\endgroup#1\@@endlink}%
\providecommand \@sanitize@url [0]{\catcode `\\12\catcode `\$12\catcode
  `\&12\catcode `\#12\catcode `\^12\catcode `\_12\catcode `\%12\relax}%
\providecommand \@@startlink[1]{}%
\providecommand \@@endlink[0]{}%
\providecommand \url  [0]{\begingroup\@sanitize@url \@url }%
\providecommand \@url [1]{\endgroup\@href {#1}{\urlprefix }}%
\providecommand \urlprefix  [0]{URL }%
\providecommand \Eprint [0]{\href }%
\providecommand \doibase [0]{http://dx.doi.org/}%
\providecommand \selectlanguage [0]{\@gobble}%
\providecommand \bibinfo  [0]{\@secondoftwo}%
\providecommand \bibfield  [0]{\@secondoftwo}%
\providecommand \translation [1]{[#1]}%
\providecommand \BibitemOpen [0]{}%
\providecommand \bibitemStop [0]{}%
\providecommand \bibitemNoStop [0]{.\EOS\space}%
\providecommand \EOS [0]{\spacefactor3000\relax}%
\providecommand \BibitemShut  [1]{\csname bibitem#1\endcsname}%
\let\auto@bib@innerbib\@empty
\bibitem [{\citenamefont {Polkovnikov}\ \emph {et~al.}(2011)\citenamefont
  {Polkovnikov}, \citenamefont {Sengupta}, \citenamefont {Silva},\ and\
  \citenamefont {Vengalattore}}]{Polkovnikov2011}%
  \BibitemOpen
  \bibfield  {author} {\bibinfo {author} {\bibfnamefont {A.}~\bibnamefont
  {Polkovnikov}}, \bibinfo {author} {\bibfnamefont {K.}~\bibnamefont
  {Sengupta}}, \bibinfo {author} {\bibfnamefont {A.}~\bibnamefont {Silva}}, \
  and\ \bibinfo {author} {\bibfnamefont {M.}~\bibnamefont {Vengalattore}},\
  }\bibfield  {title} {\enquote {\bibinfo {title} {Colloquium: Nonequilibrium
  dynamics of closed interacting quantum systems},}\ }\href {\doibase
  10.1103/RevModPhys.83.863} {\bibfield  {journal} {\bibinfo  {journal} {Rev.
  Mod. Phys.}\ }\textbf {\bibinfo {volume} {83}},\ \bibinfo {pages} {863--883}
  (\bibinfo {year} {2011})}\BibitemShut {NoStop}%
\bibitem [{\citenamefont {Lacaze}\ \emph {et~al.}(2001)\citenamefont {Lacaze},
  \citenamefont {Lallemand}, \citenamefont {Pomeau},\ and\ \citenamefont
  {Rica}}]{Lacaze2001}%
  \BibitemOpen
  \bibfield  {author} {\bibinfo {author} {\bibfnamefont {R.}~\bibnamefont
  {Lacaze}}, \bibinfo {author} {\bibfnamefont {P.}~\bibnamefont {Lallemand}},
  \bibinfo {author} {\bibfnamefont {Y.}~\bibnamefont {Pomeau}}, \ and\ \bibinfo
  {author} {\bibfnamefont {S.}~\bibnamefont {Rica}},\ }\bibfield  {title}
  {\enquote {\bibinfo {title} {Dynamical formation of a {B}ose-{E}instein
  condensate},}\ }\href {\doibase
  https://doi.org/10.1016/S0167-2789(01)00211-1} {\bibfield  {journal}
  {\bibinfo  {journal} {Physica D: Nonlinear Phenomena}\ }\textbf {\bibinfo
  {volume} {152-153}},\ \bibinfo {pages} {779 -- 786} (\bibinfo {year}
  {2001})},\ \bibinfo {note} {advances in Nonlinear Mathematics and Science: A
  Special Issue to Honor Vladimir Zakharov}\BibitemShut {NoStop}%
\bibitem [{\citenamefont {Connaughton}\ \emph {et~al.}(2005)\citenamefont
  {Connaughton}, \citenamefont {Josserand}, \citenamefont {Picozzi},
  \citenamefont {Pomeau},\ and\ \citenamefont {Rica}}]{Connaughton2005}%
  \BibitemOpen
  \bibfield  {author} {\bibinfo {author} {\bibfnamefont {C.}~\bibnamefont
  {Connaughton}}, \bibinfo {author} {\bibfnamefont {C.}~\bibnamefont
  {Josserand}}, \bibinfo {author} {\bibfnamefont {A.}~\bibnamefont {Picozzi}},
  \bibinfo {author} {\bibfnamefont {Y.}~\bibnamefont {Pomeau}}, \ and\ \bibinfo
  {author} {\bibfnamefont {S.}~\bibnamefont {Rica}},\ }\bibfield  {title}
  {\enquote {\bibinfo {title} {Condensation of classical nonlinear waves},}\
  }\href {\doibase 10.1103/PhysRevLett.95.263901} {\bibfield  {journal}
  {\bibinfo  {journal} {Phys. Rev. Lett.}\ }\textbf {\bibinfo {volume} {95}},\
  \bibinfo {pages} {263901} (\bibinfo {year} {2005})}\BibitemShut {NoStop}%
\bibitem [{\citenamefont {Sun}\ \emph {et~al.}(2012)\citenamefont {Sun},
  \citenamefont {Jia}, \citenamefont {Barsi}, \citenamefont {Rica},
  \citenamefont {Picozzi},\ and\ \citenamefont {Fleischer}}]{Sun12}%
  \BibitemOpen
  \bibfield  {author} {\bibinfo {author} {\bibfnamefont {C.}~\bibnamefont
  {Sun}}, \bibinfo {author} {\bibfnamefont {S.}~\bibnamefont {Jia}}, \bibinfo
  {author} {\bibfnamefont {C.}~\bibnamefont {Barsi}}, \bibinfo {author}
  {\bibfnamefont {S.}~\bibnamefont {Rica}}, \bibinfo {author} {\bibfnamefont
  {A.}~\bibnamefont {Picozzi}}, \ and\ \bibinfo {author} {\bibfnamefont
  {J.~W.}\ \bibnamefont {Fleischer}},\ }\bibfield  {title} {\enquote {\bibinfo
  {title} {Observation of the kinetic condensation of classical waves},}\
  }\href {\doibase 10.1038/nphys2278} {\bibfield  {journal} {\bibinfo
  {journal} {Nature Physics}\ }\textbf {\bibinfo {volume} {8}},\ \bibinfo
  {pages} {470--474} (\bibinfo {year} {2012})}\BibitemShut {NoStop}%
\bibitem [{\citenamefont {Gring}\ \emph {et~al.}(2012)\citenamefont {Gring},
  \citenamefont {Kuhnert}, \citenamefont {Langen}, \citenamefont {Kitagawa},
  \citenamefont {Rauer}, \citenamefont {Schreitl}, \citenamefont {Mazets},
  \citenamefont {Smith}, \citenamefont {Demler},\ and\ \citenamefont
  {Schmiedmayer}}]{Gring2012}%
  \BibitemOpen
  \bibfield  {author} {\bibinfo {author} {\bibfnamefont {M.}~\bibnamefont
  {Gring}}, \bibinfo {author} {\bibfnamefont {M.}~\bibnamefont {Kuhnert}},
  \bibinfo {author} {\bibfnamefont {T.}~\bibnamefont {Langen}}, \bibinfo
  {author} {\bibfnamefont {T.}~\bibnamefont {Kitagawa}}, \bibinfo {author}
  {\bibfnamefont {B.}~\bibnamefont {Rauer}}, \bibinfo {author} {\bibfnamefont
  {M.}~\bibnamefont {Schreitl}}, \bibinfo {author} {\bibfnamefont
  {I.}~\bibnamefont {Mazets}}, \bibinfo {author} {\bibfnamefont {D.~Adu}\
  \bibnamefont {Smith}}, \bibinfo {author} {\bibfnamefont {E.}~\bibnamefont
  {Demler}}, \ and\ \bibinfo {author} {\bibfnamefont {J.}~\bibnamefont
  {Schmiedmayer}},\ }\bibfield  {title} {\enquote {\bibinfo {title} {Relaxation
  and prethermalization in an isolated quantum system},}\ }\href {\doibase
  10.1126/science.1224953} {\bibfield  {journal} {\bibinfo  {journal}
  {Science}\ }\textbf {\bibinfo {volume} {337}},\ \bibinfo {pages} {1318--1322}
  (\bibinfo {year} {2012})}\BibitemShut {NoStop}%
\bibitem [{\citenamefont {Eigen}\ \emph {et~al.}(2018)\citenamefont {Eigen},
  \citenamefont {Glidden}, \citenamefont {Lopes}, \citenamefont {Cornell},
  \citenamefont {Smith},\ and\ \citenamefont {Hadzibabic}}]{Eigen18}%
  \BibitemOpen
  \bibfield  {author} {\bibinfo {author} {\bibfnamefont {C.}~\bibnamefont
  {Eigen}}, \bibinfo {author} {\bibfnamefont {J.~A.~P.}\ \bibnamefont
  {Glidden}}, \bibinfo {author} {\bibfnamefont {R.}~\bibnamefont {Lopes}},
  \bibinfo {author} {\bibfnamefont {E.~A.}\ \bibnamefont {Cornell}}, \bibinfo
  {author} {\bibfnamefont {R.~P.}\ \bibnamefont {Smith}}, \ and\ \bibinfo
  {author} {\bibfnamefont {Z.}~\bibnamefont {Hadzibabic}},\ }\bibfield  {title}
  {\enquote {\bibinfo {title} {Universal prethermal dynamics of {B}ose gases
  quenched to unitarity},}\ }\href {\doibase 10.1038/s41586-018-0674-1}
  {\bibfield  {journal} {\bibinfo  {journal} {Nature}\ }\textbf {\bibinfo
  {volume} {563}},\ \bibinfo {pages} {221--224} (\bibinfo {year}
  {2018})}\BibitemShut {NoStop}%
\bibitem [{\citenamefont {Langen}\ \emph {et~al.}(2015)\citenamefont {Langen},
  \citenamefont {Erne}, \citenamefont {Geiger}, \citenamefont {Rauer},
  \citenamefont {Schweigler}, \citenamefont {Kuhnert}, \citenamefont
  {Rohringer}, \citenamefont {Mazets}, \citenamefont {Gasenzer},\ and\
  \citenamefont {Schmiedmayer}}]{Langen2015}%
  \BibitemOpen
  \bibfield  {author} {\bibinfo {author} {\bibfnamefont {T.}~\bibnamefont
  {Langen}}, \bibinfo {author} {\bibfnamefont {S.}~\bibnamefont {Erne}},
  \bibinfo {author} {\bibfnamefont {R.}~\bibnamefont {Geiger}}, \bibinfo
  {author} {\bibfnamefont {B.}~\bibnamefont {Rauer}}, \bibinfo {author}
  {\bibfnamefont {T.}~\bibnamefont {Schweigler}}, \bibinfo {author}
  {\bibfnamefont {M.}~\bibnamefont {Kuhnert}}, \bibinfo {author} {\bibfnamefont
  {W.}~\bibnamefont {Rohringer}}, \bibinfo {author} {\bibfnamefont {I.~E.}\
  \bibnamefont {Mazets}}, \bibinfo {author} {\bibfnamefont {T.}~\bibnamefont
  {Gasenzer}}, \ and\ \bibinfo {author} {\bibfnamefont {J.}~\bibnamefont
  {Schmiedmayer}},\ }\bibfield  {title} {\enquote {\bibinfo {title}
  {Experimental observation of a generalized {G}ibbs ensemble},}\ }\href
  {\doibase 10.1126/science.1257026} {\bibfield  {journal} {\bibinfo  {journal}
  {Science}\ }\textbf {\bibinfo {volume} {348}},\ \bibinfo {pages} {207--211}
  (\bibinfo {year} {2015})}\BibitemShut {NoStop}%
\bibitem [{\citenamefont {Nicklas}\ \emph {et~al.}(2015)\citenamefont
  {Nicklas}, \citenamefont {Karl}, \citenamefont {H\"ofer}, \citenamefont
  {Johnson}, \citenamefont {Muessel}, \citenamefont {Strobel}, \citenamefont
  {Tomkovi\ifmmode~\check{c}\else \v{c}\fi{}}, \citenamefont {Gasenzer},\ and\
  \citenamefont {Oberthaler}}]{Nicklas2015}%
  \BibitemOpen
  \bibfield  {author} {\bibinfo {author} {\bibfnamefont {E.}~\bibnamefont
  {Nicklas}}, \bibinfo {author} {\bibfnamefont {M.}~\bibnamefont {Karl}},
  \bibinfo {author} {\bibfnamefont {M.}~\bibnamefont {H\"ofer}}, \bibinfo
  {author} {\bibfnamefont {A.}~\bibnamefont {Johnson}}, \bibinfo {author}
  {\bibfnamefont {W.}~\bibnamefont {Muessel}}, \bibinfo {author} {\bibfnamefont
  {H.}~\bibnamefont {Strobel}}, \bibinfo {author} {\bibfnamefont
  {J.}~\bibnamefont {Tomkovi\ifmmode~\check{c}\else \v{c}\fi{}}}, \bibinfo
  {author} {\bibfnamefont {T.}~\bibnamefont {Gasenzer}}, \ and\ \bibinfo
  {author} {\bibfnamefont {M.~K.}\ \bibnamefont {Oberthaler}},\ }\bibfield
  {title} {\enquote {\bibinfo {title} {Observation of scaling in the dynamics
  of a strongly quenched quantum gas},}\ }\href {\doibase
  10.1103/PhysRevLett.115.245301} {\bibfield  {journal} {\bibinfo  {journal}
  {Phys. Rev. Lett.}\ }\textbf {\bibinfo {volume} {115}},\ \bibinfo {pages}
  {245301} (\bibinfo {year} {2015})}\BibitemShut {NoStop}%
\bibitem [{\citenamefont {Navon}\ \emph {et~al.}(2015)\citenamefont {Navon},
  \citenamefont {Gaunt}, \citenamefont {Smith},\ and\ \citenamefont
  {Hadzibabic}}]{Navon2015}%
  \BibitemOpen
  \bibfield  {author} {\bibinfo {author} {\bibfnamefont {N.}~\bibnamefont
  {Navon}}, \bibinfo {author} {\bibfnamefont {A.~L.}\ \bibnamefont {Gaunt}},
  \bibinfo {author} {\bibfnamefont {R.~P.}\ \bibnamefont {Smith}}, \ and\
  \bibinfo {author} {\bibfnamefont {Z.}~\bibnamefont {Hadzibabic}},\ }\bibfield
   {title} {\enquote {\bibinfo {title} {Critical dynamics of spontaneous
  symmetry breaking in a homogeneous {B}ose gas},}\ }\href {\doibase
  10.1126/science.1258676} {\bibfield  {journal} {\bibinfo  {journal}
  {Science}\ }\textbf {\bibinfo {volume} {347}},\ \bibinfo {pages} {167--170}
  (\bibinfo {year} {2015})}\BibitemShut {NoStop}%
\bibitem [{\citenamefont {Berges}\ \emph {et~al.}(2008)\citenamefont {Berges},
  \citenamefont {Rothkopf},\ and\ \citenamefont {Schmidt}}]{Berges2008}%
  \BibitemOpen
  \bibfield  {author} {\bibinfo {author} {\bibfnamefont {J.}~\bibnamefont
  {Berges}}, \bibinfo {author} {\bibfnamefont {A.}~\bibnamefont {Rothkopf}}, \
  and\ \bibinfo {author} {\bibfnamefont {J.}~\bibnamefont {Schmidt}},\
  }\bibfield  {title} {\enquote {\bibinfo {title} {Nonthermal fixed points:
  Effective weak coupling for strongly correlated systems far from
  equilibrium},}\ }\href {\doibase 10.1103/PhysRevLett.101.041603} {\bibfield
  {journal} {\bibinfo  {journal} {Phys. Rev. Lett.}\ }\textbf {\bibinfo
  {volume} {101}},\ \bibinfo {pages} {041603} (\bibinfo {year}
  {2008})}\BibitemShut {NoStop}%
\bibitem [{\citenamefont {Pi\~neiro Orioli}\ \emph {et~al.}(2015)\citenamefont
  {Pi\~neiro Orioli}, \citenamefont {Boguslavski},\ and\ \citenamefont
  {Berges}}]{Orioli2015}%
  \BibitemOpen
  \bibfield  {author} {\bibinfo {author} {\bibfnamefont {A.}~\bibnamefont
  {Pi\~neiro Orioli}}, \bibinfo {author} {\bibfnamefont {K.}~\bibnamefont
  {Boguslavski}}, \ and\ \bibinfo {author} {\bibfnamefont {J.}~\bibnamefont
  {Berges}},\ }\bibfield  {title} {\enquote {\bibinfo {title} {Universal
  self-similar dynamics of relativistic and nonrelativistic field theories near
  nonthermal fixed points},}\ }\href {\doibase 10.1103/PhysRevD.92.025041}
  {\bibfield  {journal} {\bibinfo  {journal} {Phys. Rev. D}\ }\textbf {\bibinfo
  {volume} {92}},\ \bibinfo {pages} {025041} (\bibinfo {year}
  {2015})}\BibitemShut {NoStop}%
\bibitem [{\citenamefont {Erne}\ \emph {et~al.}(2018)\citenamefont {Erne},
  \citenamefont {B\"ucker}, \citenamefont {Gasenzer}, \citenamefont {Berges},\
  and\ \citenamefont {Schmiedmayer}}]{Erne2018}%
  \BibitemOpen
  \bibfield  {author} {\bibinfo {author} {\bibfnamefont {S.}~\bibnamefont
  {Erne}}, \bibinfo {author} {\bibfnamefont {R.}~\bibnamefont {B\"ucker}},
  \bibinfo {author} {\bibfnamefont {T.}~\bibnamefont {Gasenzer}}, \bibinfo
  {author} {\bibfnamefont {J.}~\bibnamefont {Berges}}, \ and\ \bibinfo {author}
  {\bibfnamefont {J.}~\bibnamefont {Schmiedmayer}},\ }\bibfield  {title}
  {\enquote {\bibinfo {title} {Universal dynamics in an isolated
  one-dimensional {B}ose gas far from equilibrium},}\ }\href {\doibase
  10.1038/s41586-018-0667-0} {\bibfield  {journal} {\bibinfo  {journal}
  {Nature}\ }\textbf {\bibinfo {volume} {563}},\ \bibinfo {pages} {225--229}
  (\bibinfo {year} {2018})}\BibitemShut {NoStop}%
\bibitem [{\citenamefont {Pr\"ufer}\ \emph {et~al.}(2018)\citenamefont
  {Pr\"ufer}, \citenamefont {Kunkel}, \citenamefont {Strobel}, \citenamefont
  {Lannig}, \citenamefont {Linnemann}, \citenamefont {Schmied}, \citenamefont
  {Berges}, \citenamefont {Gasenzer},\ and\ \citenamefont
  {Oberthaler}}]{Oberthaler2018}%
  \BibitemOpen
  \bibfield  {author} {\bibinfo {author} {\bibfnamefont {M.}~\bibnamefont
  {Pr\"ufer}}, \bibinfo {author} {\bibfnamefont {P.}~\bibnamefont {Kunkel}},
  \bibinfo {author} {\bibfnamefont {H.}~\bibnamefont {Strobel}}, \bibinfo
  {author} {\bibfnamefont {S.}~\bibnamefont {Lannig}}, \bibinfo {author}
  {\bibfnamefont {D.}~\bibnamefont {Linnemann}}, \bibinfo {author}
  {\bibfnamefont {C.-M.}\ \bibnamefont {Schmied}}, \bibinfo {author}
  {\bibfnamefont {J.}~\bibnamefont {Berges}}, \bibinfo {author} {\bibfnamefont
  {T.}~\bibnamefont {Gasenzer}}, \ and\ \bibinfo {author} {\bibfnamefont
  {M.~K.}\ \bibnamefont {Oberthaler}},\ }\bibfield  {title} {\enquote {\bibinfo
  {title} {Observation of universal dynamics in a spinor {B}ose gas far from
  equilibrium},}\ }\href {\doibase 10.1038/s41586-018-0659-0} {\bibfield
  {journal} {\bibinfo  {journal} {Nature}\ }\textbf {\bibinfo {volume} {563}},\
  \bibinfo {pages} {217--220} (\bibinfo {year} {2018})}\BibitemShut {NoStop}%
\bibitem [{\citenamefont {Huse}\ \emph {et~al.}(2014)\citenamefont {Huse},
  \citenamefont {Nandkishore},\ and\ \citenamefont {Oganesyan}}]{Huse14}%
  \BibitemOpen
  \bibfield  {author} {\bibinfo {author} {\bibfnamefont {D.~A.}\ \bibnamefont
  {Huse}}, \bibinfo {author} {\bibfnamefont {R.}~\bibnamefont {Nandkishore}}, \
  and\ \bibinfo {author} {\bibfnamefont {V.}~\bibnamefont {Oganesyan}},\
  }\bibfield  {title} {\enquote {\bibinfo {title} {Phenomenology of fully
  many-body-localized systems},}\ }\href {\doibase 10.1103/PhysRevB.90.174202}
  {\bibfield  {journal} {\bibinfo  {journal} {Phys. Rev. B}\ }\textbf {\bibinfo
  {volume} {90}},\ \bibinfo {pages} {174202} (\bibinfo {year}
  {2014})}\BibitemShut {NoStop}%
\bibitem [{\citenamefont {Alet}\ and\ \citenamefont
  {Laflorencie}(2018)}]{Alet18}%
  \BibitemOpen
  \bibfield  {author} {\bibinfo {author} {\bibfnamefont {F.}~\bibnamefont
  {Alet}}\ and\ \bibinfo {author} {\bibfnamefont {N.}~\bibnamefont
  {Laflorencie}},\ }\bibfield  {title} {\enquote {\bibinfo {title} {Many-body
  localization: An introduction and selected topics},}\ }\href {\doibase
  10.1016/j.crhy.2018.03.003} {\bibfield  {journal} {\bibinfo  {journal}
  {Comptes Rendus Physique}\ }\textbf {\bibinfo {volume} {19}},\ \bibinfo
  {pages} {498 -- 525} (\bibinfo {year} {2018})}\BibitemShut {NoStop}%
\bibitem [{\citenamefont {Basko}\ \emph {et~al.}(2006)\citenamefont {Basko},
  \citenamefont {Aleiner},\ and\ \citenamefont {Altshuler}}]{basko2006}%
  \BibitemOpen
  \bibfield  {author} {\bibinfo {author} {\bibfnamefont {D.~M.}\ \bibnamefont
  {Basko}}, \bibinfo {author} {\bibfnamefont {I.~L.}\ \bibnamefont {Aleiner}},
  \ and\ \bibinfo {author} {\bibfnamefont {B.~L.}\ \bibnamefont {Altshuler}},\
  }\bibfield  {title} {\enquote {\bibinfo {title} {Metal--insulator transition
  in a weakly interacting many-electron system with localized single-particle
  states},}\ }\href {\doibase 10.1016/j.aop.2005.11.014} {\bibfield  {journal}
  {\bibinfo  {journal} {Ann. Phys. (N.Y.)}\ }\textbf {\bibinfo {volume}
  {321}},\ \bibinfo {pages} {1126--1205} (\bibinfo {year} {2006})}\BibitemShut
  {NoStop}%
\bibitem [{\citenamefont {Mulansky}\ \emph {et~al.}(2009)\citenamefont
  {Mulansky}, \citenamefont {Ahnert}, \citenamefont {Pikovsky},\ and\
  \citenamefont {Shepelyansky}}]{Mulansky2009}%
  \BibitemOpen
  \bibfield  {author} {\bibinfo {author} {\bibfnamefont {M.}~\bibnamefont
  {Mulansky}}, \bibinfo {author} {\bibfnamefont {K.}~\bibnamefont {Ahnert}},
  \bibinfo {author} {\bibfnamefont {A.}~\bibnamefont {Pikovsky}}, \ and\
  \bibinfo {author} {\bibfnamefont {D.~L.}\ \bibnamefont {Shepelyansky}},\
  }\bibfield  {title} {\enquote {\bibinfo {title} {Dynamical thermalization of
  disordered nonlinear lattices},}\ }\href {\doibase
  10.1103/PhysRevE.80.056212} {\bibfield  {journal} {\bibinfo  {journal} {Phys.
  Rev. E}\ }\textbf {\bibinfo {volume} {80}},\ \bibinfo {pages} {056212}
  (\bibinfo {year} {2009})}\BibitemShut {NoStop}%
\bibitem [{\citenamefont {Basko}(2011)}]{Basko2011}%
  \BibitemOpen
  \bibfield  {author} {\bibinfo {author} {\bibfnamefont {D.~M.}\ \bibnamefont
  {Basko}},\ }\bibfield  {title} {\enquote {\bibinfo {title} {Weak chaos in the
  disordered nonlinear schr\"odinger chain: Destruction of {A}nderson
  localization by {A}rnold diffusion},}\ }\href {\doibase
  https://doi.org/10.1016/j.aop.2011.02.004} {\bibfield  {journal} {\bibinfo
  {journal} {Annals of Physics}\ }\textbf {\bibinfo {volume} {326}},\ \bibinfo
  {pages} {1577 -- 1655} (\bibinfo {year} {2011})},\ \bibinfo {note} {july 2011
  Special Issue}\BibitemShut {NoStop}%
\bibitem [{\citenamefont {Pikovsky}\ and\ \citenamefont
  {Shepelyansky}(2008)}]{pikovsky2008}%
  \BibitemOpen
  \bibfield  {author} {\bibinfo {author} {\bibfnamefont {A.~S.}\ \bibnamefont
  {Pikovsky}}\ and\ \bibinfo {author} {\bibfnamefont {D.~L.}\ \bibnamefont
  {Shepelyansky}},\ }\bibfield  {title} {\enquote {\bibinfo {title}
  {Destruction of {A}nderson localization by a weak nonlinearity},}\ }\href
  {\doibase 10.1103/PhysRevLett.100.094101} {\bibfield  {journal} {\bibinfo
  {journal} {Phys. Rev. Lett.}\ }\textbf {\bibinfo {volume} {100}},\ \bibinfo
  {pages} {094101} (\bibinfo {year} {2008})}\BibitemShut {NoStop}%
\bibitem [{\citenamefont {Skokos}\ \emph {et~al.}(2009)\citenamefont {Skokos},
  \citenamefont {Krimer}, \citenamefont {Komineas},\ and\ \citenamefont
  {Flach}}]{skokos2009}%
  \BibitemOpen
  \bibfield  {author} {\bibinfo {author} {\bibfnamefont {Ch.}\ \bibnamefont
  {Skokos}}, \bibinfo {author} {\bibfnamefont {D.~O.}\ \bibnamefont {Krimer}},
  \bibinfo {author} {\bibfnamefont {S.}~\bibnamefont {Komineas}}, \ and\
  \bibinfo {author} {\bibfnamefont {S.}~\bibnamefont {Flach}},\ }\bibfield
  {title} {\enquote {\bibinfo {title} {Delocalization of wave packets in
  disordered nonlinear chains},}\ }\href {\doibase 10.1103/PhysRevE.79.056211}
  {\bibfield  {journal} {\bibinfo  {journal} {Phys. Rev. E}\ }\textbf {\bibinfo
  {volume} {79}},\ \bibinfo {pages} {056211} (\bibinfo {year}
  {2009})}\BibitemShut {NoStop}%
\bibitem [{\citenamefont {Flach}\ \emph {et~al.}(2009)\citenamefont {Flach},
  \citenamefont {Krimer},\ and\ \citenamefont {Skokos}}]{flach2009}%
  \BibitemOpen
  \bibfield  {author} {\bibinfo {author} {\bibfnamefont {S.}~\bibnamefont
  {Flach}}, \bibinfo {author} {\bibfnamefont {D.~O.}\ \bibnamefont {Krimer}}, \
  and\ \bibinfo {author} {\bibfnamefont {Ch.}\ \bibnamefont {Skokos}},\
  }\bibfield  {title} {\enquote {\bibinfo {title} {Universal spreading of wave
  packets in disordered nonlinear systems},}\ }\href {\doibase
  10.1103/PhysRevLett.102.024101} {\bibfield  {journal} {\bibinfo  {journal}
  {Phys. Rev. Lett.}\ }\textbf {\bibinfo {volume} {102}},\ \bibinfo {pages}
  {024101} (\bibinfo {year} {2009})}\BibitemShut {NoStop}%
\bibitem [{\citenamefont {Cherroret}\ \emph {et~al.}(2014)\citenamefont
  {Cherroret}, \citenamefont {Vermersch}, \citenamefont {Garreau},\ and\
  \citenamefont {Delande}}]{Cherroret2014}%
  \BibitemOpen
  \bibfield  {author} {\bibinfo {author} {\bibfnamefont {N.}~\bibnamefont
  {Cherroret}}, \bibinfo {author} {\bibfnamefont {B.}~\bibnamefont
  {Vermersch}}, \bibinfo {author} {\bibfnamefont {J.~C.}\ \bibnamefont
  {Garreau}}, \ and\ \bibinfo {author} {\bibfnamefont {D.}~\bibnamefont
  {Delande}},\ }\bibfield  {title} {\enquote {\bibinfo {title} {How nonlinear
  interactions challenge the three-dimensional {A}nderson transition},}\ }\href
  {\doibase 10.1103/PhysRevLett.112.170603} {\bibfield  {journal} {\bibinfo
  {journal} {Phys. Rev. Lett.}\ }\textbf {\bibinfo {volume} {112}},\ \bibinfo
  {pages} {170603} (\bibinfo {year} {2014})}\BibitemShut {NoStop}%
\bibitem [{\citenamefont {Cherroret}(2016)}]{Cherroret2016}%
  \BibitemOpen
  \bibfield  {author} {\bibinfo {author} {\bibfnamefont {N.}~\bibnamefont
  {Cherroret}},\ }\bibfield  {title} {\enquote {\bibinfo {title} {A
  self-consistent theory of localization in nonlinear random media},}\ }\href
  {\doibase 10.1088/0953-8984/29/2/024002} {\bibfield  {journal} {\bibinfo
  {journal} {Journal of Physics: Condensed Matter}\ }\textbf {\bibinfo {volume}
  {29}},\ \bibinfo {pages} {024002} (\bibinfo {year} {2016})}\BibitemShut
  {NoStop}%
\bibitem [{\citenamefont {Vakulchyk}\ \emph {et~al.}(2019)\citenamefont
  {Vakulchyk}, \citenamefont {Fistul},\ and\ \citenamefont
  {Flach}}]{vakulchyk2018}%
  \BibitemOpen
  \bibfield  {author} {\bibinfo {author} {\bibfnamefont {I.}~\bibnamefont
  {Vakulchyk}}, \bibinfo {author} {\bibfnamefont {M.~V.}\ \bibnamefont
  {Fistul}}, \ and\ \bibinfo {author} {\bibfnamefont {S.}~\bibnamefont
  {Flach}},\ }\bibfield  {title} {\enquote {\bibinfo {title} {Wave packet
  spreading with disordered nonlinear discrete-time quantum walks},}\ }\href
  {\doibase 10.1103/PhysRevLett.122.040501} {\bibfield  {journal} {\bibinfo
  {journal} {Phys. Rev. Lett.}\ }\textbf {\bibinfo {volume} {122}},\ \bibinfo
  {pages} {040501} (\bibinfo {year} {2019})}\BibitemShut {NoStop}%
\bibitem [{\citenamefont {Billy}\ \emph {et~al.}(2008)\citenamefont {Billy},
  \citenamefont {Josse}, \citenamefont {Zuo}, \citenamefont {Bernard},
  \citenamefont {Hambrecht}, \citenamefont {Lugan}, \citenamefont {Cl\'ement},
  \citenamefont {Sanchez-Palencia}, \citenamefont {Bouyer},\ and\ \citenamefont
  {Aspect}}]{Billy2008}%
  \BibitemOpen
  \bibfield  {author} {\bibinfo {author} {\bibfnamefont {J.}~\bibnamefont
  {Billy}}, \bibinfo {author} {\bibfnamefont {V.}~\bibnamefont {Josse}},
  \bibinfo {author} {\bibfnamefont {Z.}~\bibnamefont {Zuo}}, \bibinfo {author}
  {\bibfnamefont {A.}~\bibnamefont {Bernard}}, \bibinfo {author} {\bibfnamefont
  {B.}~\bibnamefont {Hambrecht}}, \bibinfo {author} {\bibfnamefont
  {P.}~\bibnamefont {Lugan}}, \bibinfo {author} {\bibfnamefont
  {D.}~\bibnamefont {Cl\'ement}}, \bibinfo {author} {\bibfnamefont
  {L.}~\bibnamefont {Sanchez-Palencia}}, \bibinfo {author} {\bibfnamefont
  {P.}~\bibnamefont {Bouyer}}, \ and\ \bibinfo {author} {\bibfnamefont
  {A.}~\bibnamefont {Aspect}},\ }\bibfield  {title} {\enquote {\bibinfo {title}
  {Direct observation of {A}nderson localization of matter waves in a
  controlled disorder},}\ }\href {\doibase 10.1038/nature07000} {\bibfield
  {journal} {\bibinfo  {journal} {Nature}\ }\textbf {\bibinfo {volume} {453}},\
  \bibinfo {pages} {891--894} (\bibinfo {year} {2008})}\BibitemShut {NoStop}%
\bibitem [{\citenamefont {Roati}\ \emph {et~al.}(2008)\citenamefont {Roati},
  \citenamefont {D'Errico}, \citenamefont {Fallani}, \citenamefont {Fattori},
  \citenamefont {Fort}, \citenamefont {Zaccanti}, \citenamefont {Modugno},
  \citenamefont {Modugno},\ and\ \citenamefont {Inguscio}}]{Roati2008}%
  \BibitemOpen
  \bibfield  {author} {\bibinfo {author} {\bibfnamefont {G.}~\bibnamefont
  {Roati}}, \bibinfo {author} {\bibfnamefont {C.}~\bibnamefont {D'Errico}},
  \bibinfo {author} {\bibfnamefont {L.}~\bibnamefont {Fallani}}, \bibinfo
  {author} {\bibfnamefont {M.}~\bibnamefont {Fattori}}, \bibinfo {author}
  {\bibfnamefont {C.}~\bibnamefont {Fort}}, \bibinfo {author} {\bibfnamefont
  {M.}~\bibnamefont {Zaccanti}}, \bibinfo {author} {\bibfnamefont
  {G.}~\bibnamefont {Modugno}}, \bibinfo {author} {\bibfnamefont
  {M.}~\bibnamefont {Modugno}}, \ and\ \bibinfo {author} {\bibfnamefont
  {M.}~\bibnamefont {Inguscio}},\ }\bibfield  {title} {\enquote {\bibinfo
  {title} {{A}nderson localization of a non-interacting {B}ose-{E}instein
  condensate},}\ }\href {\doibase 10.1038/nature07071} {\bibfield  {journal}
  {\bibinfo  {journal} {Nature}\ }\textbf {\bibinfo {volume} {453}},\ \bibinfo
  {pages} {895--898} (\bibinfo {year} {2008})}\BibitemShut {NoStop}%
\bibitem [{\citenamefont {Jendrzejewski}\ \emph {et~al.}(2012)\citenamefont
  {Jendrzejewski}, \citenamefont {M\"uller}, \citenamefont {Richard},
  \citenamefont {Date}, \citenamefont {Plisson}, \citenamefont {Bouyer},
  \citenamefont {Aspect},\ and\ \citenamefont {Josse}}]{Jendrzejewski2012}%
  \BibitemOpen
  \bibfield  {author} {\bibinfo {author} {\bibfnamefont {F.}~\bibnamefont
  {Jendrzejewski}}, \bibinfo {author} {\bibfnamefont {K.}~\bibnamefont
  {M\"uller}}, \bibinfo {author} {\bibfnamefont {J.}~\bibnamefont {Richard}},
  \bibinfo {author} {\bibfnamefont {A.}~\bibnamefont {Date}}, \bibinfo {author}
  {\bibfnamefont {T.}~\bibnamefont {Plisson}}, \bibinfo {author} {\bibfnamefont
  {P.}~\bibnamefont {Bouyer}}, \bibinfo {author} {\bibfnamefont
  {A.}~\bibnamefont {Aspect}}, \ and\ \bibinfo {author} {\bibfnamefont
  {V.}~\bibnamefont {Josse}},\ }\bibfield  {title} {\enquote {\bibinfo {title}
  {Coherent backscattering of ultracold atoms},}\ }\href {\doibase
  10.1103/PhysRevLett.109.195302} {\bibfield  {journal} {\bibinfo  {journal}
  {Phys. Rev. Lett.}\ }\textbf {\bibinfo {volume} {109}},\ \bibinfo {pages}
  {195302} (\bibinfo {year} {2012})}\BibitemShut {NoStop}%
\bibitem [{\citenamefont {Labeyrie}\ \emph {et~al.}(2012)\citenamefont
  {Labeyrie}, \citenamefont {Karpiuk}, \citenamefont {Schaff}, \citenamefont
  {Gr{\'{e}}maud}, \citenamefont {Miniatura},\ and\ \citenamefont
  {Delande}}]{Labeyrie2012}%
  \BibitemOpen
  \bibfield  {author} {\bibinfo {author} {\bibfnamefont {G.}~\bibnamefont
  {Labeyrie}}, \bibinfo {author} {\bibfnamefont {T.}~\bibnamefont {Karpiuk}},
  \bibinfo {author} {\bibfnamefont {J.-F.}\ \bibnamefont {Schaff}}, \bibinfo
  {author} {\bibfnamefont {B.}~\bibnamefont {Gr{\'{e}}maud}}, \bibinfo {author}
  {\bibfnamefont {C.}~\bibnamefont {Miniatura}}, \ and\ \bibinfo {author}
  {\bibfnamefont {D.}~\bibnamefont {Delande}},\ }\bibfield  {title} {\enquote
  {\bibinfo {title} {Enhanced backscattering of a dilute {B}ose-{E}instein
  condensate},}\ }\href {\doibase 10.1209/0295-5075/100/66001} {\bibfield
  {journal} {\bibinfo  {journal} {{EPL} (Europhysics Letters)}\ }\textbf
  {\bibinfo {volume} {100}},\ \bibinfo {pages} {66001} (\bibinfo {year}
  {2012})}\BibitemShut {NoStop}%
\bibitem [{\citenamefont {Hainaut}\ \emph {et~al.}(2017)\citenamefont
  {Hainaut}, \citenamefont {Manai}, \citenamefont {Chicireanu}, \citenamefont
  {Cl\'ement}, \citenamefont {Zemmouri}, \citenamefont {Garreau}, \citenamefont
  {Szriftgiser}, \citenamefont {Lemari\'e}, \citenamefont {Cherroret},\ and\
  \citenamefont {Delande}}]{Hainaut2017}%
  \BibitemOpen
  \bibfield  {author} {\bibinfo {author} {\bibfnamefont {C.}~\bibnamefont
  {Hainaut}}, \bibinfo {author} {\bibfnamefont {I.}~\bibnamefont {Manai}},
  \bibinfo {author} {\bibfnamefont {R.}~\bibnamefont {Chicireanu}}, \bibinfo
  {author} {\bibfnamefont {J.-F.}\ \bibnamefont {Cl\'ement}}, \bibinfo {author}
  {\bibfnamefont {S.}~\bibnamefont {Zemmouri}}, \bibinfo {author}
  {\bibfnamefont {J.~C.}\ \bibnamefont {Garreau}}, \bibinfo {author}
  {\bibfnamefont {P.}~\bibnamefont {Szriftgiser}}, \bibinfo {author}
  {\bibfnamefont {G.}~\bibnamefont {Lemari\'e}}, \bibinfo {author}
  {\bibfnamefont {N.}~\bibnamefont {Cherroret}}, \ and\ \bibinfo {author}
  {\bibfnamefont {D.}~\bibnamefont {Delande}},\ }\bibfield  {title} {\enquote
  {\bibinfo {title} {Return to the origin as a probe of atomic phase
  coherence},}\ }\href {\doibase 10.1103/PhysRevLett.118.184101} {\bibfield
  {journal} {\bibinfo  {journal} {Phys. Rev. Lett.}\ }\textbf {\bibinfo
  {volume} {118}},\ \bibinfo {pages} {184101} (\bibinfo {year}
  {2017})}\BibitemShut {NoStop}%
\bibitem [{\citenamefont {M\"uller}\ \emph {et~al.}(2015)\citenamefont
  {M\"uller}, \citenamefont {Richard}, \citenamefont {Volchkov}, \citenamefont
  {Denechaud}, \citenamefont {Bouyer}, \citenamefont {Aspect},\ and\
  \citenamefont {Josse}}]{Josse2015}%
  \BibitemOpen
  \bibfield  {author} {\bibinfo {author} {\bibfnamefont {K.}~\bibnamefont
  {M\"uller}}, \bibinfo {author} {\bibfnamefont {J.}~\bibnamefont {Richard}},
  \bibinfo {author} {\bibfnamefont {V.~V.}\ \bibnamefont {Volchkov}}, \bibinfo
  {author} {\bibfnamefont {V.}~\bibnamefont {Denechaud}}, \bibinfo {author}
  {\bibfnamefont {P.}~\bibnamefont {Bouyer}}, \bibinfo {author} {\bibfnamefont
  {A.}~\bibnamefont {Aspect}}, \ and\ \bibinfo {author} {\bibfnamefont
  {V.}~\bibnamefont {Josse}},\ }\bibfield  {title} {\enquote {\bibinfo {title}
  {Suppression and revival of weak localization through control of
  time-reversal symmetry},}\ }\href {\doibase 10.1103/PhysRevLett.114.205301}
  {\bibfield  {journal} {\bibinfo  {journal} {Phys. Rev. Lett.}\ }\textbf
  {\bibinfo {volume} {114}},\ \bibinfo {pages} {205301} (\bibinfo {year}
  {2015})}\BibitemShut {NoStop}%
\bibitem [{\citenamefont {Cherroret}\ \emph {et~al.}(2012)\citenamefont
  {Cherroret}, \citenamefont {Karpiuk}, \citenamefont {M\"uller}, \citenamefont
  {Gr\'emaud},\ and\ \citenamefont {Miniatura}}]{Cherroret2011}%
  \BibitemOpen
  \bibfield  {author} {\bibinfo {author} {\bibfnamefont {N.}~\bibnamefont
  {Cherroret}}, \bibinfo {author} {\bibfnamefont {T.}~\bibnamefont {Karpiuk}},
  \bibinfo {author} {\bibfnamefont {C.~A.}\ \bibnamefont {M\"uller}}, \bibinfo
  {author} {\bibfnamefont {B.}~\bibnamefont {Gr\'emaud}}, \ and\ \bibinfo
  {author} {\bibfnamefont {C.}~\bibnamefont {Miniatura}},\ }\bibfield  {title}
  {\enquote {\bibinfo {title} {Coherent backscattering of ultracold matter
  waves: Momentum space signatures},}\ }\href {\doibase
  10.1103/PhysRevA.85.011604} {\bibfield  {journal} {\bibinfo  {journal} {Phys.
  Rev. A}\ }\textbf {\bibinfo {volume} {85}},\ \bibinfo {pages} {011604}
  (\bibinfo {year} {2012})}\BibitemShut {NoStop}%
\bibitem [{\citenamefont {Karpiuk}\ \emph {et~al.}(2012)\citenamefont
  {Karpiuk}, \citenamefont {Cherroret}, \citenamefont {Lee}, \citenamefont
  {Gr\'emaud}, \citenamefont {M\"uller},\ and\ \citenamefont
  {Miniatura}}]{Karpiuk2012}%
  \BibitemOpen
  \bibfield  {author} {\bibinfo {author} {\bibfnamefont {T.}~\bibnamefont
  {Karpiuk}}, \bibinfo {author} {\bibfnamefont {N.}~\bibnamefont {Cherroret}},
  \bibinfo {author} {\bibfnamefont {K.~L.}\ \bibnamefont {Lee}}, \bibinfo
  {author} {\bibfnamefont {B.}~\bibnamefont {Gr\'emaud}}, \bibinfo {author}
  {\bibfnamefont {C.~A.}\ \bibnamefont {M\"uller}}, \ and\ \bibinfo {author}
  {\bibfnamefont {C.}~\bibnamefont {Miniatura}},\ }\bibfield  {title} {\enquote
  {\bibinfo {title} {Coherent forward scattering peak induced by {A}nderson
  localization},}\ }\href {\doibase 10.1103/PhysRevLett.109.190601} {\bibfield
  {journal} {\bibinfo  {journal} {Phys. Rev. Lett.}\ }\textbf {\bibinfo
  {volume} {109}},\ \bibinfo {pages} {190601} (\bibinfo {year}
  {2012})}\BibitemShut {NoStop}%
\bibitem [{\citenamefont {Micklitz}\ \emph {et~al.}(2015)\citenamefont
  {Micklitz}, \citenamefont {M\"uller},\ and\ \citenamefont
  {Altland}}]{Micklitz2015}%
  \BibitemOpen
  \bibfield  {author} {\bibinfo {author} {\bibfnamefont {T.}~\bibnamefont
  {Micklitz}}, \bibinfo {author} {\bibfnamefont {C.~A.}\ \bibnamefont
  {M\"uller}}, \ and\ \bibinfo {author} {\bibfnamefont {A.}~\bibnamefont
  {Altland}},\ }\bibfield  {title} {\enquote {\bibinfo {title} {Echo
  spectroscopy of {A}nderson localization},}\ }\href {\doibase
  10.1103/PhysRevB.91.064203} {\bibfield  {journal} {\bibinfo  {journal} {Phys.
  Rev. B}\ }\textbf {\bibinfo {volume} {91}},\ \bibinfo {pages} {064203}
  (\bibinfo {year} {2015})}\BibitemShut {NoStop}%
\bibitem [{\citenamefont {Cherroret}\ \emph {et~al.}(2015)\citenamefont
  {Cherroret}, \citenamefont {Karpiuk}, \citenamefont {Gr\'emaud},\ and\
  \citenamefont {Miniatura}}]{Cherroret2015}%
  \BibitemOpen
  \bibfield  {author} {\bibinfo {author} {\bibfnamefont {N.}~\bibnamefont
  {Cherroret}}, \bibinfo {author} {\bibfnamefont {T.}~\bibnamefont {Karpiuk}},
  \bibinfo {author} {\bibfnamefont {B.}~\bibnamefont {Gr\'emaud}}, \ and\
  \bibinfo {author} {\bibfnamefont {C.}~\bibnamefont {Miniatura}},\ }\bibfield
  {title} {\enquote {\bibinfo {title} {Thermalization of matter waves in
  speckle potentials},}\ }\href {\doibase 10.1103/PhysRevA.92.063614}
  {\bibfield  {journal} {\bibinfo  {journal} {Phys. Rev. A}\ }\textbf {\bibinfo
  {volume} {92}},\ \bibinfo {pages} {063614} (\bibinfo {year}
  {2015})}\BibitemShut {NoStop}%
\bibitem [{\citenamefont {Wellens}\ and\ \citenamefont
  {Gr\'emaud}(2008)}]{Wellens2008}%
  \BibitemOpen
  \bibfield  {author} {\bibinfo {author} {\bibfnamefont {T.}~\bibnamefont
  {Wellens}}\ and\ \bibinfo {author} {\bibfnamefont {B.}~\bibnamefont
  {Gr\'emaud}},\ }\bibfield  {title} {\enquote {\bibinfo {title} {Nonlinear
  {coherent} {transport} of {waves} in {disordered} {media}},}\ }\href
  {\doibase 10.1103/PhysRevLett.100.033902} {\bibfield  {journal} {\bibinfo
  {journal} {Phys. Rev. Lett.}\ }\textbf {\bibinfo {volume} {100}} (\bibinfo
  {year} {2008}),\ 10.1103/PhysRevLett.100.033902}\BibitemShut {NoStop}%
\bibitem [{\citenamefont {Wellens}(2009)}]{Wellens2009}%
  \BibitemOpen
  \bibfield  {author} {\bibinfo {author} {\bibfnamefont {T.}~\bibnamefont
  {Wellens}},\ }\bibfield  {title} {\enquote {\bibinfo {title} {Nonlinear
  coherent backscattering},}\ }\href {\doibase 10.1007/s00340-009-3454-7}
  {\bibfield  {journal} {\bibinfo  {journal} {Applied Physics B}\ }\textbf
  {\bibinfo {volume} {95}},\ \bibinfo {pages} {189} (\bibinfo {year}
  {2009})}\BibitemShut {NoStop}%
\bibitem [{\citenamefont {Geiger}\ \emph {et~al.}(2012)\citenamefont {Geiger},
  \citenamefont {Wellens},\ and\ \citenamefont {Buchleitner}}]{Geiger2012}%
  \BibitemOpen
  \bibfield  {author} {\bibinfo {author} {\bibfnamefont {T.}~\bibnamefont
  {Geiger}}, \bibinfo {author} {\bibfnamefont {T.}~\bibnamefont {Wellens}}, \
  and\ \bibinfo {author} {\bibfnamefont {A.}~\bibnamefont {Buchleitner}},\
  }\bibfield  {title} {\enquote {\bibinfo {title} {Inelastic multiple
  scattering of interacting bosons in weak random potentials},}\ }\href
  {\doibase 10.1103/PhysRevLett.109.030601} {\bibfield  {journal} {\bibinfo
  {journal} {Phys. Rev. Lett.}\ }\textbf {\bibinfo {volume} {109}},\ \bibinfo
  {pages} {030601} (\bibinfo {year} {2012})}\BibitemShut {NoStop}%
\bibitem [{\citenamefont {Geiger}\ \emph {et~al.}(2013)\citenamefont {Geiger},
  \citenamefont {Buchleitner},\ and\ \citenamefont {Wellens}}]{Geiger2013}%
  \BibitemOpen
  \bibfield  {author} {\bibinfo {author} {\bibfnamefont {T.}~\bibnamefont
  {Geiger}}, \bibinfo {author} {\bibfnamefont {A.}~\bibnamefont {Buchleitner}},
  \ and\ \bibinfo {author} {\bibfnamefont {T.}~\bibnamefont {Wellens}},\
  }\bibfield  {title} {\enquote {\bibinfo {title} {Microscopic scattering
  theory for interacting bosons in weak random potentials},}\ }\href {\doibase
  10.1088/1367-2630/15/11/115015} {\bibfield  {journal} {\bibinfo  {journal}
  {New Journal of Physics}\ }\textbf {\bibinfo {volume} {15}},\ \bibinfo
  {pages} {115015} (\bibinfo {year} {2013})}\BibitemShut {NoStop}%
\bibitem [{\citenamefont {Hartung}\ \emph {et~al.}(2008)\citenamefont
  {Hartung}, \citenamefont {Wellens}, \citenamefont {M\"uller}, \citenamefont
  {Richter},\ and\ \citenamefont {Schlagheck}}]{Hartung2008}%
  \BibitemOpen
  \bibfield  {author} {\bibinfo {author} {\bibfnamefont {M.}~\bibnamefont
  {Hartung}}, \bibinfo {author} {\bibfnamefont {T.}~\bibnamefont {Wellens}},
  \bibinfo {author} {\bibfnamefont {C.~A.}\ \bibnamefont {M\"uller}}, \bibinfo
  {author} {\bibfnamefont {K.}~\bibnamefont {Richter}}, \ and\ \bibinfo
  {author} {\bibfnamefont {P.}~\bibnamefont {Schlagheck}},\ }\bibfield  {title}
  {\enquote {\bibinfo {title} {Coherent backscattering of {B}ose-{E}instein
  condensates in two-dimensional disorder potentials},}\ }\href {\doibase
  10.1103/PhysRevLett.101.020603} {\bibfield  {journal} {\bibinfo  {journal}
  {Phys. Rev. Lett.}\ }\textbf {\bibinfo {volume} {101}},\ \bibinfo {pages}
  {020603} (\bibinfo {year} {2008})}\BibitemShut {NoStop}%
\bibitem [{\citenamefont {Tal-Ezer}\ and\ \citenamefont
  {Kosloff}(1984)}]{Tal-Ezer84}%
  \BibitemOpen
  \bibfield  {author} {\bibinfo {author} {\bibfnamefont {H.}~\bibnamefont
  {Tal-Ezer}}\ and\ \bibinfo {author} {\bibfnamefont {R.}~\bibnamefont
  {Kosloff}},\ }\bibfield  {title} {\enquote {\bibinfo {title} {An accurate and
  efficient scheme for propagating the time dependent schr\"odinger
  equation},}\ }\href {\doibase https://doi.org/10.1063/1.448136} {\bibfield
  {journal} {\bibinfo  {journal} {J. Chem. Phys.}\ }\textbf {\bibinfo {volume}
  {81}},\ \bibinfo {pages} {3967} (\bibinfo {year} {1984})}\BibitemShut
  {NoStop}%
\bibitem [{\citenamefont {Leforestier}\ \emph {et~al.}(1991)\citenamefont
  {Leforestier}, \citenamefont {Bisseling}, \citenamefont {Cerjan},
  \citenamefont {Feit}, \citenamefont {Friesner}, \citenamefont {Guldberg},
  \citenamefont {Hammerich}, \citenamefont {Jolicard}, \citenamefont
  {Karrlein}, \citenamefont {Meyer}, \citenamefont {Lipkin}, \citenamefont
  {Roncero},\ and\ \citenamefont {Kosloff}}]{Cheby91}%
  \BibitemOpen
  \bibfield  {author} {\bibinfo {author} {\bibfnamefont {C.}~\bibnamefont
  {Leforestier}}, \bibinfo {author} {\bibfnamefont {R.~H.}\ \bibnamefont
  {Bisseling}}, \bibinfo {author} {\bibfnamefont {C.}~\bibnamefont {Cerjan}},
  \bibinfo {author} {\bibfnamefont {M.~D.}\ \bibnamefont {Feit}}, \bibinfo
  {author} {\bibfnamefont {R.}~\bibnamefont {Friesner}}, \bibinfo {author}
  {\bibfnamefont {A.}~\bibnamefont {Guldberg}}, \bibinfo {author}
  {\bibfnamefont {A.}~\bibnamefont {Hammerich}}, \bibinfo {author}
  {\bibfnamefont {G.}~\bibnamefont {Jolicard}}, \bibinfo {author}
  {\bibfnamefont {W.}~\bibnamefont {Karrlein}}, \bibinfo {author}
  {\bibfnamefont {H.-D.}\ \bibnamefont {Meyer}}, \bibinfo {author}
  {\bibfnamefont {N.}~\bibnamefont {Lipkin}}, \bibinfo {author} {\bibfnamefont
  {O.}~\bibnamefont {Roncero}}, \ and\ \bibinfo {author} {\bibfnamefont
  {R.}~\bibnamefont {Kosloff}},\ }\bibfield  {title} {\enquote {\bibinfo
  {title} {A comparison of different propagation schemes for the time dependent
  schr\"odinger equation},}\ }\href {\doibase
  https://doi.org/10.1016/0021-9991(91)90137-A} {\bibfield  {journal} {\bibinfo
   {journal} {J. Comput. Phys.}\ }\textbf {\bibinfo {volume} {94}},\ \bibinfo
  {pages} {59} (\bibinfo {year} {1991})}\BibitemShut {NoStop}%
\bibitem [{\citenamefont {Roche}\ and\ \citenamefont
  {Mayou}(1997)}]{roche1997conductivity}%
  \BibitemOpen
  \bibfield  {author} {\bibinfo {author} {\bibfnamefont {S.}~\bibnamefont
  {Roche}}\ and\ \bibinfo {author} {\bibfnamefont {D.}~\bibnamefont {Mayou}},\
  }\bibfield  {title} {\enquote {\bibinfo {title} {Conductivity of
  quasiperiodic systems: A numerical study},}\ }\href {\doibase
  10.1103/PhysRevLett.79.2518} {\bibfield  {journal} {\bibinfo  {journal}
  {Phys. Rev. Lett.}\ }\textbf {\bibinfo {volume} {79}},\ \bibinfo {pages}
  {2518--2521} (\bibinfo {year} {1997})}\BibitemShut {NoStop}%
\bibitem [{\citenamefont {Fehske}\ \emph {et~al.}(2009)\citenamefont {Fehske},
  \citenamefont {Schleede}, \citenamefont {Schubert}, \citenamefont {Wellein},
  \citenamefont {Filinov},\ and\ \citenamefont {Bishop}}]{fehske2009numerical}%
  \BibitemOpen
  \bibfield  {author} {\bibinfo {author} {\bibfnamefont {H.}~\bibnamefont
  {Fehske}}, \bibinfo {author} {\bibfnamefont {J.}~\bibnamefont {Schleede}},
  \bibinfo {author} {\bibfnamefont {G.}~\bibnamefont {Schubert}}, \bibinfo
  {author} {\bibfnamefont {G.}~\bibnamefont {Wellein}}, \bibinfo {author}
  {\bibfnamefont {V.~S.}\ \bibnamefont {Filinov}}, \ and\ \bibinfo {author}
  {\bibfnamefont {A.~R.}\ \bibnamefont {Bishop}},\ }\bibfield  {title}
  {\enquote {\bibinfo {title} {Numerical approaches to time evolution of
  complex quantum systems},}\ }\href {\doibase
  https://doi.org/10.1016/j.physleta.2009.04.022} {\bibfield  {journal}
  {\bibinfo  {journal} {Physics Letters A}\ }\textbf {\bibinfo {volume}
  {373}},\ \bibinfo {pages} {2182 -- 2188} (\bibinfo {year}
  {2009})}\BibitemShut {NoStop}%
\bibitem [{\citenamefont {Kuhn}\ \emph {et~al.}(2005)\citenamefont {Kuhn},
  \citenamefont {Miniatura}, \citenamefont {Delande}, \citenamefont
  {Sigwarth},\ and\ \citenamefont {M\"uller}}]{Kuhn05}%
  \BibitemOpen
  \bibfield  {author} {\bibinfo {author} {\bibfnamefont {R.~C.}\ \bibnamefont
  {Kuhn}}, \bibinfo {author} {\bibfnamefont {C.}~\bibnamefont {Miniatura}},
  \bibinfo {author} {\bibfnamefont {D.}~\bibnamefont {Delande}}, \bibinfo
  {author} {\bibfnamefont {O.}~\bibnamefont {Sigwarth}}, \ and\ \bibinfo
  {author} {\bibfnamefont {C.~A.}\ \bibnamefont {M\"uller}},\ }\bibfield
  {title} {\enquote {\bibinfo {title} {Localization of matter waves in
  two-dimensional disordered optical potentials},}\ }\href {\doibase
  10.1103/PhysRevLett.95.250403} {\bibfield  {journal} {\bibinfo  {journal}
  {Phys. Rev. Lett.}\ }\textbf {\bibinfo {volume} {95}},\ \bibinfo {pages}
  {250403} (\bibinfo {year} {2005})}\BibitemShut {NoStop}%
\bibitem [{\citenamefont {Kuhn}\ \emph {et~al.}(2007)\citenamefont {Kuhn},
  \citenamefont {Sigwarth}, \citenamefont {Miniatura}, \citenamefont
  {Delande},\ and\ \citenamefont {M\"uller}}]{Kuhn07}%
  \BibitemOpen
  \bibfield  {author} {\bibinfo {author} {\bibfnamefont {R.~C.}\ \bibnamefont
  {Kuhn}}, \bibinfo {author} {\bibfnamefont {O.}~\bibnamefont {Sigwarth}},
  \bibinfo {author} {\bibfnamefont {C.}~\bibnamefont {Miniatura}}, \bibinfo
  {author} {\bibfnamefont {D.}~\bibnamefont {Delande}}, \ and\ \bibinfo
  {author} {\bibfnamefont {C.~A.}\ \bibnamefont {M\"uller}},\ }\bibfield
  {title} {\enquote {\bibinfo {title} {Coherent matter wave transport in
  speckle potentials},}\ }\href {\doibase 10.1088/1367-2630/9/6/161} {\bibfield
   {journal} {\bibinfo  {journal} {New Journal of Physics}\ }\textbf {\bibinfo
  {volume} {9}},\ \bibinfo {pages} {161--161} (\bibinfo {year}
  {2007})}\BibitemShut {NoStop}%
\bibitem [{Note1()}]{Note1}%
  \BibitemOpen
  \bibinfo {note} {Note that our definition of the occupation number
  $f_{\epsilon }$ does not make it dimensionless, see Eq. (\ref
  {eq:normalization_cond}), unlike what is more frequently encountered in the
  literature. We adopt this convention because when $g\not =0$ it makes the
  natural energy scale for interactions $g\rho _0\equiv gN/V$ explicitly appear
  in all our expressions, rather than $gN$ and $V$ separately. A dimensionless
  $f_{\epsilon }$ could nevertheless be chosen by normalizing the energy
  distribution $\nu _{\epsilon }f_{\epsilon }$ to $\rho _0$ rather than to
  unity in Eq. (\ref {eq:normalization_cond}).}\BibitemShut {Stop}%
\bibitem [{\citenamefont {Plisson}\ \emph {et~al.}(2013)\citenamefont
  {Plisson}, \citenamefont {Bourdel},\ and\ \citenamefont
  {M\"uller}}]{Plisson2013}%
  \BibitemOpen
  \bibfield  {author} {\bibinfo {author} {\bibfnamefont {T.}~\bibnamefont
  {Plisson}}, \bibinfo {author} {\bibfnamefont {T.}~\bibnamefont {Bourdel}}, \
  and\ \bibinfo {author} {\bibfnamefont {C.~A.}\ \bibnamefont {M\"uller}},\
  }\bibfield  {title} {\enquote {\bibinfo {title} {Momentum isotropisation in
  random potentials},}\ }\href {\doibase
  https://doi.org/10.1140/epjst/e2013-01756-8} {\bibfield  {journal} {\bibinfo
  {journal} {The European Physical Journal Special Topics}\ }\textbf {\bibinfo
  {volume} {217}},\ \bibinfo {pages} {79} (\bibinfo {year} {2013})}\BibitemShut
  {NoStop}%
\bibitem [{\citenamefont {Cherroret}\ and\ \citenamefont
  {Wellens}(2011)}]{CherroretWellens11}%
  \BibitemOpen
  \bibfield  {author} {\bibinfo {author} {\bibfnamefont {N.}~\bibnamefont
  {Cherroret}}\ and\ \bibinfo {author} {\bibfnamefont {T.}~\bibnamefont
  {Wellens}},\ }\bibfield  {title} {\enquote {\bibinfo {title} {Fokker-{P}lanck
  equation for transport of wave packets in nonlinear disordered media},}\
  }\href {\doibase 10.1103/PhysRevE.84.021114} {\bibfield  {journal} {\bibinfo
  {journal} {Phys. Rev. E}\ }\textbf {\bibinfo {volume} {84}},\ \bibinfo
  {pages} {021114} (\bibinfo {year} {2011})}\BibitemShut {NoStop}%
\bibitem [{\citenamefont {Schwiete}\ and\ \citenamefont
  {Finkel'stein}(2010)}]{Schwiete2010}%
  \BibitemOpen
  \bibfield  {author} {\bibinfo {author} {\bibfnamefont {G.}~\bibnamefont
  {Schwiete}}\ and\ \bibinfo {author} {\bibfnamefont {A.~M.}\ \bibnamefont
  {Finkel'stein}},\ }\bibfield  {title} {\enquote {\bibinfo {title} {Nonlinear
  wave-packet dynamics in a disordered medium},}\ }\href {\doibase
  10.1103/PhysRevLett.104.103904} {\bibfield  {journal} {\bibinfo  {journal}
  {Phys. Rev. Lett.}\ }\textbf {\bibinfo {volume} {104}},\ \bibinfo {pages}
  {103904} (\bibinfo {year} {2010})}\BibitemShut {NoStop}%
\bibitem [{\citenamefont {Schwiete}\ and\ \citenamefont
  {Finkel'stein}(2013{\natexlab{a}})}]{Schwiete2013a}%
  \BibitemOpen
  \bibfield  {author} {\bibinfo {author} {\bibfnamefont {G.}~\bibnamefont
  {Schwiete}}\ and\ \bibinfo {author} {\bibfnamefont {A.~M.}\ \bibnamefont
  {Finkel'stein}},\ }\bibfield  {title} {\enquote {\bibinfo {title} {Effective
  theory for the propagation of a wave packet in a disordered and nonlinear
  medium},}\ }\href {\doibase 10.1103/PhysRevA.87.043636} {\bibfield  {journal}
  {\bibinfo  {journal} {Phys. Rev. A}\ }\textbf {\bibinfo {volume} {87}},\
  \bibinfo {pages} {043636} (\bibinfo {year} {2013}{\natexlab{a}})}\BibitemShut
  {NoStop}%
\bibitem [{\citenamefont {Akkermans}\ and\ \citenamefont
  {Montambaux}(2007)}]{akkermans2007mesoscopic}%
  \BibitemOpen
  \bibfield  {author} {\bibinfo {author} {\bibfnamefont {E.}~\bibnamefont
  {Akkermans}}\ and\ \bibinfo {author} {\bibfnamefont {G.}~\bibnamefont
  {Montambaux}},\ }\href@noop {} {\emph {\bibinfo {title} {Mesoscopic physics
  of electrons and photons}}}\ (\bibinfo  {publisher} {Cambridge University
  Press},\ \bibinfo {year} {2007})\BibitemShut {NoStop}%
\bibitem [{\citenamefont {Pitaevskii}\ and\ \citenamefont
  {Stringari}(2016)}]{pitaevskii2016bose}%
  \BibitemOpen
  \bibfield  {author} {\bibinfo {author} {\bibfnamefont {L.}~\bibnamefont
  {Pitaevskii}}\ and\ \bibinfo {author} {\bibfnamefont {S.}~\bibnamefont
  {Stringari}},\ }\href@noop {} {\emph {\bibinfo {title} {{B}ose-{E}instein
  condensation and superfluidity}}},\ Vol.\ \bibinfo {volume} {164}\ (\bibinfo
  {publisher} {Oxford University Press},\ \bibinfo {year} {2016})\BibitemShut
  {NoStop}%
\bibitem [{\citenamefont {Schwiete}\ and\ \citenamefont
  {Finkel'stein}(2013{\natexlab{b}})}]{Schwiete2013b}%
  \BibitemOpen
  \bibfield  {author} {\bibinfo {author} {\bibfnamefont {G.}~\bibnamefont
  {Schwiete}}\ and\ \bibinfo {author} {\bibfnamefont {A.~M.}\ \bibnamefont
  {Finkel'stein}},\ }\bibfield  {title} {\enquote {\bibinfo {title} {Kinetics
  of the disordered {B}ose gas with collisions},}\ }\href {\doibase
  10.1103/PhysRevA.88.053611} {\bibfield  {journal} {\bibinfo  {journal} {Phys.
  Rev. A}\ }\textbf {\bibinfo {volume} {88}},\ \bibinfo {pages} {053611}
  (\bibinfo {year} {2013}{\natexlab{b}})}\BibitemShut {NoStop}%
\bibitem [{\citenamefont {Griffin}\ and\ \citenamefont
  {Zaremba}(2009)}]{Griffin09}%
  \BibitemOpen
  \bibfield  {author} {\bibinfo {author} {\bibfnamefont {T.~N.~A.}\
  \bibnamefont {Griffin}}\ and\ \bibinfo {author} {\bibfnamefont
  {E.}~\bibnamefont {Zaremba}},\ }\href@noop {} {\emph {\bibinfo {title}
  {{B}ose-Condensed Gases at Finite Temperatures}}}\ (\bibinfo  {publisher}
  {Cambridge University Press, Cambridge},\ \bibinfo {year} {2009})\BibitemShut
  {NoStop}%
\bibitem [{\citenamefont {Altshuler}\ and\ \citenamefont
  {Aronov}(1985)}]{Altshuler1985}%
  \BibitemOpen
  \bibfield  {author} {\bibinfo {author} {\bibfnamefont {B.~L.}\ \bibnamefont
  {Altshuler}}\ and\ \bibinfo {author} {\bibfnamefont {A.~G.}\ \bibnamefont
  {Aronov}},\ }\href@noop {} {\emph {\bibinfo {title} {Electron-electron
  interaction in disordered conductors, in Electron-electron Interactions in
  Disordered Systems}}},\ edited by\ \bibinfo {editor} {\bibfnamefont {A.~L.}\
  \bibnamefont {Efros}}\ and\ \bibinfo {editor} {\bibfnamefont
  {M.}~\bibnamefont {Pollak}}\ (\bibinfo {year} {1985})\ p.~\bibinfo {pages}
  {1}\BibitemShut {NoStop}%
\bibitem [{\citenamefont {Ashcroft}\ and\ \citenamefont
  {Mermin}(1976)}]{Ashcroft76}%
  \BibitemOpen
  \bibfield  {author} {\bibinfo {author} {\bibfnamefont {N.}~\bibnamefont
  {Ashcroft}}\ and\ \bibinfo {author} {\bibfnamefont {D.}~\bibnamefont
  {Mermin}},\ }\href@noop {} {\emph {\bibinfo {title} {Solid State Physics}}}\
  (\bibinfo  {publisher} {Philadelphia, PA: Saunders College},\ \bibinfo {year}
  {1976})\BibitemShut {NoStop}%
\bibitem [{Note2()}]{Note2}%
  \BibitemOpen
  \bibinfo {note} {As is well known, adding CBS to the diffusive contribution
  leads to a breakdown of normalization when $g=0$. This is also the case when
  $g\not =0$, as seen from the fact that $\DOTSI \intop \ilimits@ {\protect \rm
  d}\epsilon \protect \tmspace +\thinmuskip {.1667em} \nu _\epsilon f_\epsilon
  ^C(t)\not =0$. Restoring normalization requires a fine account of weak
  localization corrections \cite {Knothe13}, which is beyond the scope of this
  paper.}\BibitemShut {Stop}%
\bibitem [{\citenamefont {Trappe}\ \emph {et~al.}(2015)\citenamefont {Trappe},
  \citenamefont {Delande},\ and\ \citenamefont {M\"uller}}]{Trappe15}%
  \BibitemOpen
  \bibfield  {author} {\bibinfo {author} {\bibfnamefont {M.~I.}\ \bibnamefont
  {Trappe}}, \bibinfo {author} {\bibfnamefont {D.}~\bibnamefont {Delande}}, \
  and\ \bibinfo {author} {\bibfnamefont {C.~A.}\ \bibnamefont {M\"uller}},\
  }\bibfield  {title} {\enquote {\bibinfo {title} {Semiclassical spectral
  function for matter waves in random potentials},}\ }\href {\doibase
  10.1088/1751-8113/48/24/245102} {\bibfield  {journal} {\bibinfo  {journal}
  {Journal of Physics A: Mathematical and Theoretical}\ }\textbf {\bibinfo
  {volume} {48}},\ \bibinfo {pages} {245102} (\bibinfo {year}
  {2015})}\BibitemShut {NoStop}%
\bibitem [{\citenamefont {Prat}\ \emph {et~al.}(2016)\citenamefont {Prat},
  \citenamefont {Cherroret},\ and\ \citenamefont {Delande}}]{Prat16}%
  \BibitemOpen
  \bibfield  {author} {\bibinfo {author} {\bibfnamefont {T.}~\bibnamefont
  {Prat}}, \bibinfo {author} {\bibfnamefont {N.}~\bibnamefont {Cherroret}}, \
  and\ \bibinfo {author} {\bibfnamefont {D.}~\bibnamefont {Delande}},\
  }\bibfield  {title} {\enquote {\bibinfo {title} {Semiclassical spectral
  function and density of states in speckle potentials},}\ }\href {\doibase
  10.1103/PhysRevA.94.022114} {\bibfield  {journal} {\bibinfo  {journal} {Phys.
  Rev. A}\ }\textbf {\bibinfo {volume} {94}},\ \bibinfo {pages} {022114}
  (\bibinfo {year} {2016})}\BibitemShut {NoStop}%
\bibitem [{\citenamefont {Volchkov}\ \emph {et~al.}(2018)\citenamefont
  {Volchkov}, \citenamefont {Pasek}, \citenamefont {Denechaud}, \citenamefont
  {Mukhtar}, \citenamefont {Aspect}, \citenamefont {Delande},\ and\
  \citenamefont {Josse}}]{Volchkov18}%
  \BibitemOpen
  \bibfield  {author} {\bibinfo {author} {\bibfnamefont {V.~V.}\ \bibnamefont
  {Volchkov}}, \bibinfo {author} {\bibfnamefont {M.}~\bibnamefont {Pasek}},
  \bibinfo {author} {\bibfnamefont {V.}~\bibnamefont {Denechaud}}, \bibinfo
  {author} {\bibfnamefont {M.}~\bibnamefont {Mukhtar}}, \bibinfo {author}
  {\bibfnamefont {A.}~\bibnamefont {Aspect}}, \bibinfo {author} {\bibfnamefont
  {D.}~\bibnamefont {Delande}}, \ and\ \bibinfo {author} {\bibfnamefont
  {V.}~\bibnamefont {Josse}},\ }\bibfield  {title} {\enquote {\bibinfo {title}
  {Measurement of spectral functions of ultracold atoms in disordered
  potentials},}\ }\href {\doibase 10.1103/PhysRevLett.120.060404} {\bibfield
  {journal} {\bibinfo  {journal} {Phys. Rev. Lett.}\ }\textbf {\bibinfo
  {volume} {120}},\ \bibinfo {pages} {060404} (\bibinfo {year}
  {2018})}\BibitemShut {NoStop}%
\bibitem [{\citenamefont {Carleo}\ \emph {et~al.}(2013)\citenamefont {Carleo},
  \citenamefont {Bo\'eris}, \citenamefont {Holzmann},\ and\ \citenamefont
  {Sanchez-Palencia}}]{Carleo2013}%
  \BibitemOpen
  \bibfield  {author} {\bibinfo {author} {\bibfnamefont {G.}~\bibnamefont
  {Carleo}}, \bibinfo {author} {\bibfnamefont {G.}~\bibnamefont {Bo\'eris}},
  \bibinfo {author} {\bibfnamefont {M.}~\bibnamefont {Holzmann}}, \ and\
  \bibinfo {author} {\bibfnamefont {L.}~\bibnamefont {Sanchez-Palencia}},\
  }\bibfield  {title} {\enquote {\bibinfo {title} {Universal superfluid
  transition and transport properties of two-dimensional dirty bosons},}\
  }\href {\doibase 10.1103/PhysRevLett.111.050406} {\bibfield  {journal}
  {\bibinfo  {journal} {Phys. Rev. Lett.}\ }\textbf {\bibinfo {volume} {111}},\
  \bibinfo {pages} {050406} (\bibinfo {year} {2013})}\BibitemShut {NoStop}%
\bibitem [{\citenamefont {Sanchez-Palencia}(2006)}]{Sanchez-Palencia06}%
  \BibitemOpen
  \bibfield  {author} {\bibinfo {author} {\bibfnamefont {L.}~\bibnamefont
  {Sanchez-Palencia}},\ }\bibfield  {title} {\enquote {\bibinfo {title}
  {Smoothing effect and delocalization of interacting {B}ose-{E}instein
  condensates in random potentials},}\ }\href {\doibase
  10.1103/PhysRevA.74.053625} {\bibfield  {journal} {\bibinfo  {journal} {Phys.
  Rev. A}\ }\textbf {\bibinfo {volume} {74}},\ \bibinfo {pages} {053625}
  (\bibinfo {year} {2006})}\BibitemShut {NoStop}%
\bibitem [{\citenamefont {Watson}(1966)}]{Watson66}%
  \BibitemOpen
  \bibfield  {author} {\bibinfo {author} {\bibfnamefont {G.~N.}\ \bibnamefont
  {Watson}},\ }\href@noop {} {\emph {\bibinfo {title} {A Treatise on the Theory
  of Bessel Functions, 2nd ed.}}}\ (\bibinfo  {publisher} {Cambridge, England:
  Cambridge University Press},\ \bibinfo {year} {1966})\BibitemShut {NoStop}%
\bibitem [{\citenamefont {Knothe}\ and\ \citenamefont
  {Wellens}(2013)}]{Knothe13}%
  \BibitemOpen
  \bibfield  {author} {\bibinfo {author} {\bibfnamefont {A.}~\bibnamefont
  {Knothe}}\ and\ \bibinfo {author} {\bibfnamefont {T.}~\bibnamefont
  {Wellens}},\ }\bibfield  {title} {\enquote {\bibinfo {title} {Flux
  conservation in coherent backscattering and weak localization of light},}\
  }\href {\doibase 10.1088/1751-8113/46/31/315101} {\bibfield  {journal}
  {\bibinfo  {journal} {Journal of Physics A: Mathematical and Theoretical}\
  }\textbf {\bibinfo {volume} {46}},\ \bibinfo {pages} {315101} (\bibinfo
  {year} {2013})}\BibitemShut {NoStop}%
\end{thebibliography}
%

\end{document}